\documentclass[12pt]{article}
\pdfoutput=1
\usepackage[nosort]{cite}
\usepackage[normalem]{ulem}

\usepackage[dvipsnames]{xcolor}
\usepackage{epsfig}
\usepackage{amsfonts}
\usepackage{amscd}
\usepackage{latexsym}
\usepackage{amsmath, amssymb}
\usepackage{verbatim}
\usepackage{color}
\usepackage{fancyhdr}
\usepackage{hyperref}
\usepackage{tikz}
\usepackage{slashed}
\usepackage{multirow}
\usetikzlibrary{calc}
\usetikzlibrary{topaths}
\usetikzlibrary{decorations}
\usetikzlibrary{decorations.pathmorphing}
\usetikzlibrary{arrows,decorations.markings,cd}
\usetikzlibrary{calc,arrows,cd,decorations.markings,snakes}
\tikzset{
->-/.style args={#1rotate#2}{decoration={markings, mark=at position #1 with {\arrow[scale=1.5,rotate = #2 ]{stealth}}}, postaction={decorate}}
}
\usetikzlibrary{shapes.geometric}
\usetikzlibrary{knots}
\usepackage{tikz-3dplot}

\usepackage{float}
\usepackage{draft}
\usepackage{graphicx, subfig}
\usepackage{mciteplus}
\usepackage{skak}
\usepackage{bbm}

\usepackage{amsthm}
\usepackage{cleveref}
\usepackage{algorithm}
\usepackage[letterpaper,left=1.1in,top=1in,right=1.1in,bottom=1.3in]{geometry}
\usepackage{makecell}
\usepackage{enumerate}
\usepackage{tablefootnote}
\usepackage{longtable}
\numberwithin{equation}{section}

\usepackage[us,hhmmss]{datetime}

\usepackage{pifont}

\usepackage[scr=euler]{mathalfa}

\usepackage{subcaption}
\usepackage[english]{babel}
\usepackage{soul}
\usepackage{upgreek}
\usepackage[scr]{rsfso}

\usepackage{tikz}

\usepackage{longtable}

\usepackage{amsmath,amssymb}

\DeclareMathAlphabet{\zapfcal}{OT1}{pzc}{m}{it}
\newcommand{\cl}{\zapfcal{l}}
\newcommand{\cs}{\zapfcal{S}}

\numberwithin{equation}{section}

\def\C{\mathcal{C}}

\def\R{\mathcal{R}}

\def\T{\mathcal{T}}

\def\Ts{{\mathbb{T}^2}}
\def\Kb{{\mathbb{K}^2}}

\def\sC{\mathsf{C}}

\def\sT{\mathsf{T}}
\def\sR{\mathsf{R}}

\def\sJ{\mathsf{J}}
\def\sL{\mathsf{L}}

\def\scT{\mathscr{T}}
\def\scG{\mathscr{G}}
\def\scK{\mathscr{K}}

\def\Aut{\mathrm{Aut}}

\colorlet{mylinkcolor}{NavyBlue}
\colorlet{mycitecolor}{Aquamarine}
\colorlet{myurlcolor}{Aquamarine}
\newcommand\myshade{90}

\hypersetup{
  linkcolor  = mylinkcolor!\myshade!black,
  citecolor  = mycitecolor!\myshade!black,
  urlcolor   = myurlcolor!\myshade!black,
  colorlinks = true,
}

\begin{document}
\begin{titlepage}

\title{Tori, Klein Bottles, and Modulo 8 Parity/Time-reversal Anomalies of 2+1d Staggered Fermions}

\author{Nathan Seiberg$^1$ and Wucheng Zhang$^2$}

 \address{${}^1$ School of Natural Sciences, Institute for Advanced Study, Princeton, NJ}
 \address{${}^2$ Department of Physics, Princeton University, Princeton, NJ}

\abstract{
\noindent We study the symmetries of lattice staggered fermions in 2+1d.  Using the symmetries, we can place the system on any sheared torus or Klein bottle.  These different backgrounds provide diagnostics of various 't Hooft anomalies associated with the crystalline symmetries. We then compare the lattice model to its continuum limit. The symmetries of the lattice system are mapped in a nontrivial way to the symmetries of the continuum theories.  Using this map, we match the 't Hooft anomalies on the lattice and the continuum. Along the way, we develop a general formalism to study Hamiltonian lattice models on nontrivial, compact, flat spaces. 
}

\end{titlepage}
\tableofcontents

\section{Introduction}
\label{sec:intro}
't Hooft anomalies lead to powerful constraints on the dynamics of complicated problems \cite{tHooft:1979rat}.  In more detail, one identifies the symmetries and their anomalies in the short-distance UV theory.  Then, a similar analysis is done for any candidate long-distance IR theory.  The matching between these UV and IR anomalies can then be used as a constraint on the candidate IR theory.

Of particular interest is the case when the UV system is on a lattice, and the IR system is in the continuum.  In this case, the match of the symmetries and their anomalies can be nontrivial.  Examples are the LSM constraints \cite{Lieb:1961fr,Affleck:1986pq,Affleck:1988nt,Oshikawa:2000lrt,Hastings:2003zx} and their interpretation as associated with anomalies  \cite{Cheng:2015kce,Jian:2017skd,Cho:2017fgz,Metlitski:2017fmd,Else:2019lft,Gioia:2021xtp,Cheng:2022sgb}.  In particular, the UV crystalline symmetries can become internal symmetries in the IR, and the UV anomalies should then matched with anomalies in IR internal symmetries. More generally, the authors of \cite{Thorngren:2016hdm} put forward an interesting proposal about anomalies of such crystalline symmetries.

In this note, we will focus on anomalies involving spatial reflection (parity) and time-reversal in 2+1d fermionic systems.\footnote{In our lattice systems, spatial reflections and time-reversal symmetries are different, and so are their anomalies.  In a relativistic system, they are the same.  In this latter context, the terms ``parity anomaly'' and ``time-reversal anomaly'' are used interchangeably.\label{timeparity}}  In the continuum, the parity anomaly first appeared in the physics literature in \cite{Niemi:1983rq,Redlich:1983kn,Redlich:1983dv,Alvarez-Gaume:1984zst,Rao:1986ba}, and has since been analyzed in more detail in many papers, culminating in \cite{Kapustin:2014dxa,Cho:2015ega,Hsieh:2015xaa,Witten:2015aba,Witten:2016cio,Wang:2016qkb,Tachikawa:2016cha, Tachikawa:2016nmo,Barkeshli:2016mew,Tata:2021jwp}. On the lattice, this and related anomalies were studied in \cite{Chen:2010zpc,doi:10.1073/pnas.1514665112,Misumi:2019jrt,Ogata:2020hry,Yao:2021cdh,Cheng:2022sgb,Seiberg:2023cdc,Yao:2023bnj,Seiberg:2024gek,Pace:2024oys,Pace:2025rfu,Kim:2025tzc}.  In particular, \cite{Seiberg:2025zqx} discussed a closely related anomaly in the Majorana chain in 1+1 dimensions.  Here, we will build on this work and extend it to 2+1 dimensions.

An important aspect of the anomaly is its order.  To define it, we consider $N_f$ identical copies of our system, and define the order as the smallest value of $N_f$ for which the anomaly is absent.  Since finding this order might not be easy, we establish lower and upper bounds on it.
\begin{itemize}
    \item A priori, the internal symmetry acts independently on each of the $N_f$ copies, but we focus only on the diagonal one.  The spacetime symmetries must act the same on all the copies.  Then, we look for a deformation of the system, possibly coupling the $N_f$ copies in a way that preserves the diagonal symmetry that gaps the system with a trivial ground state.  If we succeed, the anomaly is at most of order $N_f$. 
    \item We study the system on various spacetimes and find anomalous phases.  In this note, we will take space to be a torus or a Klein bottle and allow various twists along the Euclidean time directions.  A lower bound on the order of the anomaly is the value of $N_f$ such that all the anomalous phases vanish.
\end{itemize}
In favorable situations, the lower and upper bounds coincide, allowing us to deduce the value of the order of the anomaly.

We will focus on the simplest fermionic model, i.e., a real one-component fermion at each site with the staggered coupling \cite{Kogut:1974ag,Banks:1975gq, Banks:1976ia,Susskind:1976jm} 
\ie\label{latticeHi}
    &H=i \sum_{\vec\ell,\mu=1,2} \eta_\mu(\vec\ell) \psi_{\vec{\ell}+\hat{\mu}} \psi_{\vec{\ell}}\,,\qquad\eta_\mu(\vec\ell)=(-1)^{\sum_{\nu<\mu} \ell^\nu}=\begin{cases} 1&\mu=1\\ (-1)^{\ell_1}&\mu=2 \end{cases}\,.
\fe
(Unlike most of the literature about this model, here the fermions are real.)
The position-dependent $\eta_\mu(\vec\ell)$ are naturally interpreted as a background $\mathbb{Z}_2$ gauge field with ``$\pi$-flux'' through each plaquette. (See also more recent studies of staggered fermions in \cite{Affleck:2017ubr,Catterall:2022jky,Catterall:2024jps, Chatterjee:2024gje,Li:2024dpq,Catterall:2025vrx}.)

We will perform a detailed analysis of the symmetries of this model and their 't Hooft anomalies.  We will do it by placing the model on various flat spaces with twisted boundary conditions and identifying the projective representations of the symmetry operators.

The internal symmetry is fermion parity $\mathbb{Z}_2^F$.

The nontrivial crystalline symmetry of the model was studied in \cite{vandenDoel:1983mf,Kluberg-Stern:1983lmr, Parisi:1983rv, Golterman:1984dn, Golterman:1985dz,  Golterman:1986jf, Kilcup:1986dg}.  In particular, the two translation generators  $T_{1,2}$ do not commute
\ie
  T_1T_2=(-1)^F T_2T_1\,.
\fe

The total symmetry group $G$ of the infinite lattice also includes ${\pi \over 2}$-rotation $\C$, spatial reflection $\R$, and time-reversal $\T$.

We would like to study the theory on a compact, locally flat homogeneous space $X$.\footnote{We will refer interchangeably both to the compact continuous space and the lattice as $X$.}  Then, for 2d space, $X$ can only be a torus or a Klein bottle. To achieve this, we identify points on the infinite lattice $\bZ^2$ and specify the corresponding boundary conditions. Specifically, we pick two group elements\footnote{Here, we follow our notation on the lattice. Our notation in the continuum is $g_I\to k_I$, $G\to K$, and $\scG\to \scK$.} $g_1, g_2 \in G$ and impose the identifications
\ie
\psi_\ell=g_1 \psi_\ell g_1^{-1}=g_2 \psi_\ell g_2^{-1}\,.
\fe
These group elements generate $\scG\subset G$. Then, to obtain the flat space $X$, i.e., a torus or a Klein bottle, we need the group to be the fundamental group $\scG = \langle g_1,g_2\rangle \cong \pi_1(X)$.

Once we place the theory on a finite lattice $X$, we should find its symmetry. It is
\ie\label{globalGint}
  G_X = {N_G(\scG)\over \scG}\,,
\fe
where $N_G(\scG)$ is the normalizer of $\scG$ in $G$. (This will be discussed further in Sections \ref{sec:2+1dcont} and \ref{sec:staggeredfermions}.) Distinct models correspond to subgroups $\scG$ of $G$ that are inequivalent under conjugation.

For any one of the distinct models, we will find its global symmetry group $G_X$ \eqref{globalGint} and will identify the Hilbert space as a projective representation of $G_X$. The projective phases in this representation reflect the anomaly.  As we will see, more sophisticated backgrounds can probe stricter lower bounds of the anomaly order, and the lattice system has a modulo 8 anomaly.

We will repeat this discussion in the continuum limit.  In that limit, the lattice theory flows to a single, free, Dirac fermion $\Psi$ with an $O(2)$ internal symmetry.  This continuum system is known to have a parity/time-reversal anomaly \cite{Niemi:1983rq,Redlich:1983kn,Redlich:1983dv,Alvarez-Gaume:1984zst,Rao:1986ba, Wang:2014lca,Metlitski:2014xqa,Hsieh:2015xaa,Witten:2015aba,Witten:2016cio,Wang:2016qkb,Tachikawa:2016cha,Tachikawa:2016nmo,Barkeshli:2016mew,Tata:2021jwp}. Following the analysis of \cite{vandenDoel:1983mf,Kluberg-Stern:1983lmr, Parisi:1983rv, Golterman:1984dn, Golterman:1985dz,  Golterman:1986jf, Kilcup:1986dg}, we will map the lattice crystalline symmetry to the symmetry of the continuum theory.  This map is nontrivial because some of the continuum internal $O(2)$ symmetries emanate from the UV crystalline symmetries.  See \cite{Cheng:2022sgb} for a discussion of emanant symmetries.

Using this map of the symmetries, we will match the various twists of the lattice theory to twists of the continuum theory.  Then, we will match the unbroken global symmetry and its projective representations.  This highly nontrivial matching identifies modulo 8 anomalies of the crystalline symmetry, including spatial reflection and time-reversal, with anomalies involving parity/time-reversal and the internal symmetry in the continuum.

The rest of the paper is organized as follows. Section \ref{sec:2+1dcont} will be devoted to the 2+1d continuum Dirac fermion.  Here, we will review its parity anomaly in a manner that facilitates comparison with the lattice model.   Section \ref{sec:staggeredfermions} will be devoted to the 2+1d Majorana staggered fermion.  Here, we will study the symmetries and anomalies on the lattice under various twists. Finally, in Section \ref{sec:tocontinuum}, we will match the lattice and continuum models, including spatial twists, symmetries, and anomalies. We will conclude our discussion in Section \ref{Conclusions}, where we will summarize our results.

In the appendices, we will provide more background material and detailed derivations. In Appendix \ref{app:revlat}, we will review the modulated symmetries of the naive lattice fermions and will use them to motivate the staggered fermion model. In Appendix \ref{app:internal}, we will examine the general formula \eqref{globalGint}, and will show that it generalizes the known result, involving the centralizer, for twists in internal symmetries. In Appendix \ref{app:decode}, we will derive various corollaries of \eqref{globalGint}, and will use them to classify different twists both on the lattice and in the continuum. In Appendix \ref{app:kleinfermioncont}, we will review the symmetries and the normal modes of the continuum fermions on a Klein bottle. In Appendix \ref{app:twistedcontinuum}, we will discuss the staggered fermions with various twists, match them with the continuum models, and derive the anomalies. Finally, in Appendix \ref{app:Omegaanomaly}, we will map the anomaly of the lattice symmetry operator $\Omega=\T\C^2$ in our 2+1-dimensional problem to the time-reversal anomaly in quantum mechanics. 

\section{Continuum 2+1d Dirac fermion and parity anomaly}
\label{sec:2+1dcont}

In this section, we will review well-known facts about the parity anomaly of the 2+1d continuum Dirac fermion (see, e.g., \cite{Niemi:1983rq,Redlich:1983kn,Redlich:1983dv,Alvarez-Gaume:1984zst,Rao:1986ba,Witten:2015aba,Witten:2016cio}).  The purpose of this review is to introduce our conventions for the symmetry action, and the various twisted compactifications (on $\Ts$ and $\Kb$).  Our approach here, which is based on the symmetry algebra of the model on these spaces, follows the discussion in \cite{Delmastro:2021xox}.

\subsection{Symmetries}
We denote the massless Dirac fermion field in 2+1d continuum as two massless Majorana fermions, $\Psi_{\alpha=1,2}$. Here we suppress the Lorentz indices, and $\alpha$ is a flavor index.

We impose the global $O(2)=U(1) \rtimes \bZ_2$ symmetry, generated by $\bZ_2$ charge conjugation $\Gamma$ and the $U(1)$ generator $\sJ$, where\footnote{Here we use the conventions
\ie
\sigma_1=\left(\begin{matrix}
0 & 1 \\
1 & 0
\end{matrix}\right) \quad, \quad \sigma_2=\left(\begin{matrix}
0 & -i \\
i & 0
\end{matrix}\right) \quad, \quad \sigma_3=\left(\begin{matrix}
1 & 0 \\
0 & -1
\end{matrix}\right)\,.
\fe
}
\ie
[\sJ,\Psi_\alpha]&=(\sigma_2)^{\alpha\beta}\,\Psi_\beta\,,\\
\Gamma \Psi_\alpha \Gamma^{-1} & = (\sigma_3)^{\alpha\beta}\,\Psi_\beta\,.
\fe
The $\bZ_2$ Fermion parity symmetry is a subgroup of the $U(1)$ part, as $e^{i \pi \sJ} = (-1)^F$. They obey the group relation,  
\ie \label{eq:O2}
\Gamma^2 = 1\,,\quad e^{2\pi \sJ} = 1\,,\quad \Gamma e^{i \theta \sJ}  = e^{-i \theta \sJ}\Gamma\,.
\fe
Since we would like to match this continuum theory with a non-relativistic Hamiltonian lattice model, we will focus on the space reflection $\sR$ along axis 1, the anti-unitary time-reversal $\sT$, and the spatial rotation $e^{i\theta \sL}$.  They act on the fermions as\footnote{Here, we construct the Clifford algebra $\mathrm{Cl}_{2,1}(\mathbb{R})$ for Majorana fermions in 2+1d. The gamma matrices are $\gamma^1=\sigma_1$, $\gamma^2=-\sigma_3$, and $\gamma^0 = \gamma^1\gamma^2=i\sigma^2$.\label{ft:gammadefn}} 
\ie
\sR \Psi_\alpha(x^1,x^2,t) \sR^{-1} &= \gamma_1\Psi_\alpha(-x^1,x^2,t)\,,\\
\sT \Psi_\alpha(x^1,x^2,t) \sT^{-1} &= \gamma_0\Psi_\alpha(x^1,x^2,-t)\,,\\
e^{i\theta \sL} \Psi_\alpha(x^1,x^2,t) e^{-i\theta \sL} &= e^{\frac{1}{2}\theta  \gamma^0}\Psi_\alpha(\cos \theta\, x^1 -\sin\theta\, x^2,\sin\theta\, x^1 + \cos \theta\, x^2,t)\,,\\
\fe
and they obey the group relation
\ie
& \sR^2 = 1\,, \quad e^{2\pi i \sL} = \sT^2 = (-1)^F\,, \quad \sR e^{i\theta \sL} = e^{-i\theta \sL} \sR\,,\\
& \sR \sT = (-1)^F \sT \sR\,, \quad \sT e^{i\theta \sL} = e^{i\theta \sL} \sT\,.
\fe
This means that they form the group $\mathrm{Pin}_+(2) \times \bZ_2^\Xi$, where the $\mathrm{Pin}_+(2)$ factor is generated by $e^{i\theta \sL}$ and $\sR$, and the $\bZ_2^\Xi$ factor is generated by the anti-unitary $\Xi = \sT e^{i\pi \sL}$. Note that $\Xi$ commutes with all the other transformations here.\footnote{All unitary relativistic systems have a CRT symmetry.  (See \cite{Seiberg:2025zqx} for a recent review and references to earlier papers.)  The anti-unitary transformation $\Xi$ is not CRT because it reflects both time and the two spatial directions.}  For more details, see, e.g., \cite{Lawson:1998yr}.

These operators form the symmetry group
\ie
\frac{O(2) \times \left(\mathrm{Pin}_+(2) \times \bZ_2^{\Xi}\right)}{\bZ_2^F}\,.
\fe
Finally, we include the translation symmetry on the plane $\bR^2$, represented as
\ie
e^{i \left(\rho^1 P_1+\rho^2  P_2\right)}
\fe
where $P_i$ is the momentum operator along $x^i$-axis and this operator is the translation by $(\rho^1,\rho^2)$. The rotation, spatial reflection, and time-reversal in
$ \mathrm{Pin}_+(2) \times \bZ_2$ acts non-trivially on the momentum operators,
\ie
k \begin{pmatrix} P_1 \\ P_2 \end{pmatrix} k^{-1}
= U(k)\begin{pmatrix} P_1 \\ P_2 \end{pmatrix}\,,
\fe
where
\ie \label{eq:Ulabel}
U(\sR) =
\begin{pmatrix} -1 & 0 \\ 0 & 1 \end{pmatrix},\quad
U(\sT) =
\begin{pmatrix} -1 & 0 \\ 0 & -1 \end{pmatrix},\quad
U\left(e^{i\theta \sL}\right) =
\begin{pmatrix} \cos\theta & \sin\theta \\ -\sin\theta & \cos\theta \end{pmatrix}.
\fe
Thus, the full symmetry of interest is given by
\ie
K = \frac{O(2)\times \left(\bR^2 \rtimes (\mathrm{Pin}_+(2) \times \bZ_2^\Xi )\right)}{\bZ_2^F}\,.
\fe

To study the anomalies, we consider a system made of $N_f$ copies of the original system, which consists of a single Dirac fermion.  The invariant Lagrangian is
\ie
\mathcal L = i\sum_{A=1}^{N_f}\sum_{\alpha=1}^2 \,(\Psi_\alpha^{A})^{T}\, \gamma^0\gamma^\mu\partial_\mu \Psi_\alpha^{A}\,,
\fe
where the symmetries act diagonally on the $N_f$ species.  Next, we will determine the minimal value of $N_f$ such that the projective phases vanish for all spacetime and internal symmetry twists, and interpret this value as the lower bound of the order of the anomaly.  

Below, to avoid cluttering the equations, we suppress the $N_f$ copies and index $A$ but present the $N_f$ dependence when it is nontrivial.

\subsection{Twist by symmetries: torus and Klein bottle}
\label{sec:conttwist}

\subsubsection{Twisted boundary conditions}
\label{sec:ptid}
Here, we summarize the standard constructions of the torus and Klein bottle, with the choices of Spin or Pin structure.  See e.g., \cite{kirby1990pin,hatcher2002algebraic,Hsieh:2015xaa,Kaidi:2019tyf}.

We begin with the infinite plane $\mathbb{R}^2$ and construct compact manifolds by geometric identifications of coordinates.

Using the isometry of the problem, there are two inequivalent cases: the torus and the Klein bottle
\ie\label{simplT2K2}
&\mathbb{T}^2:\qquad (x^1,x^2) \sim (x^1+\cs^1_1,x^2) \sim (x^1+\cs^1_2, x^2+\cs^2_2)\\
&\mathbb{K}^2: \qquad (x^1,x^2) \sim (x^1+\cs^1_1,x^2) \sim (-x^1, x^2+\cs^2_2)\,.
\fe

Next, we place the Dirac fermions on these two manifolds and study the possible boundary conditions. More mathematically, this amounts to choosing a $\mathrm{Pin}_+$ structure. On the torus, there is a preferred spin structure, the odd spin structure, with
\ie \label{eq:torusbc}
\mathbb{T}^2:\qquad \Psi_\alpha(x^1,x^2) = \Psi_\alpha(x^1+\cs^1_1,x^2)\,, \quad \Psi_\alpha(x^1,x^2) = \Psi_\alpha(x^1+\cs^1_2, x^2+\cs^2_2)\,.
\fe
Other spin structures, with $\pm$ signs in these relations, are obtained by adding a $\bZ_2$ gauge field for $(-1)^F$.\footnote{Modular transformations on the torus preserve the odd spin structure and permute the three other (even) ones. See Subsection \ref{sec:equivcont}.} On the Klein bottle, we can take 
\ie  \label{eq:kbbc}
\mathbb{K}^2:\qquad \Psi_\alpha(x^1,x^2) =  \Psi_\alpha(x^1+\cs^1_1,x^2)\,, \quad \Psi_\alpha(x^1,x^2) =  \gamma^1\Psi_\alpha(-x^1, x^2+\cs^2_2)\,.
\fe
Again, other signs can be viewed as a background $\bZ_2$ gauge field for $(-1)^F$.\footnote{Unlike the torus, here, there are two preferred sign choices, the one in \eqref{eq:kbbc} and an additional minus sign in the second equation \eqref{eq:kbbc}. These two options are invariant under the allowed modular transformations, which add cycle 1 to cycle 2.  The two other sign choices (with a minus sign in the first equation in \eqref{eq:kbbc})
are exchanged under this transformation.}

\subsubsection{Group theory of the general twists}
These boundary conditions can be reformulated as follows.
The boundary conditions \eqref{eq:torusbc} and \eqref{eq:kbbc} can be recast as
\ie\label{kactiononPsi}
k_I\Psi_\alpha k_I^{-1} =\Psi_\alpha\,,
\fe
where for torus
\ie \label{eq:contT2}
k_1 = e^{i \cs^1_1 {P}_1}\,, \quad k_2 = e^{i (\cs^1_2 {P}_1 + \cs^2_2 {P}_2)}\,,
\fe
and on the Klein bottle,
\ie \label{eq:contK2}
k_1 = e^{i \cs^1_1 {P}_1}\,, \quad k_2 = e^{i \cs^1_2 {P}_1}\sR\,.
\fe
(Note that these $k_1$ and $k_2$ can be taken to be independent of the rotation elements in $K$.  Later, when we discuss the lattice counterpart of this analysis in Section \ref{sec:staggeredfermions},  this fact will not be true.)

We can verify that $k_{1,2}$ generates the fundamental groups of the manifolds,
\ie
& \langle k_1,k_2 | k_1k_2 = k_2k_1\rangle  \cong \pi_1(\Ts) \cong \bZ^2\,,\\
& \langle k_1,k_2 | k_2k_1k_2^{-1} = k_1^{-1}\rangle \cong  \pi_1(\Kb) \cong \bZ \rtimes \bZ\,,\\
\fe
and denote the generated group as $\scK = \langle k_1,k_2\rangle$. As one can recombine the boundary conditions in \eqref{eq:torusbc} or \eqref{eq:kbbc}, we have the full set of boundary conditions
\ie \label{eq:fullbc}
k \Psi_\alpha k^{-1} = \Psi_\alpha\,, \quad \forall k \in \scK\,.
\fe

Internal $O(2)$ symmetry twists can also be incorporated into $k_{1,2}$ as defects. However, the generated group $\scK$ must remain the fundamental group of the manifold, so that different combinations of the boundary conditions are consistent with each other. (See more mathematical details in e.g., \cite{hatcher2002algebraic,Ratcliffe:2006bfa}.) In the cases we study, one can first define the geometry of the manifolds and the $\mathrm{Pin}_+$ structure on it, then insert the internal symmetry twists along each cycle. Then, the internal symmetry twists should obey the relations from the fundamental group of the base manifold.

Finally, we note the important fact that the action of $\scK$ on the covering space should be free, i.e., without a fixed point.

\subsubsection{The symmetry of the twisted model}\label{symmetryoftwistedcon}

Once we place the model on a compact manifold, we would like to find the remaining symmetry. To do that, we must identify the operators $h \in K$ that preserve the conditions \eqref{eq:fullbc}, namely 
\ie\label{khpsi}
k h\Psi_\alpha h^{-1}k^{-1} = h\Psi_\alpha h^{-1}\,, \quad \forall k \in \scK\,.
\fe
Therefore, all these $h$ elements form the normalizer of $\scK$,
\ie
N_K(\scK) = \{h\in K |h k h^{-1} \in \scK, \forall k \in \scK \}\,.
\fe
Moreover, on the compact manifold, all $k \in \scK$ act trivially on the fermion fields $\Psi_\alpha$, constraining $k=1$. Thus, to obtain the faithful symmetry of the theory on the compact manifold $K_\text{compact}$, we must quotient by $\scK$,\footnote{\label{footnote:isometry}  A similar expression arises in a different, but related setup.  Consider a theory with a global symmetry $K$ and gauge a subgroup $\scK\subset K$, which acts as $\Psi \to k\Psi k^{-1} $ (compare with \eqref{kactiononPsi}).  Then, the unbroken global symmetry is \eqref{khpsi}, and it leads to the global symmetry after gauging $N_K(\scK)/\scK$.  (We ignore here the possibility that, after gauging, 't Hooft anomalies can lead to ABJ anomalies and thus break the symmetry.) However, the two problems are really different.  In the gauge theory, we restrict attention to the gauge invariant subset of the Hilbert space and keep any fields that are composed out of the fundamental field $\Psi$ and invariant under $k$. In contrast, in this paper, we set the non-gauge invariant fundamental fields $\Psi$ to zero, i.e., we impose \eqref{kactiononPsi} on the fundamental fields.}
\ie \label{eq:contsym}
K_\text{compact} = {N_K(\scK)\over\scK}\,.
\fe
In the simple case where $K$ and $\scK$ include only geometric actions, this group is the isometry of the compact space.  See, e.g., \cite{hatcher2002algebraic,Ratcliffe:2006bfa}.

The identification specified by $\scK$ is different from the internal symmetry twists. Internal symmetry twists are defects of each cycle, specified only by internal symmetries. The remaining internal symmetry of the compact manifold should commute with the twists. See more discussion in Appendix \ref{app:internal}.

Above, we obtained models on compact flat manifolds by identifying points in $\bR^2$ using $\scK$. Since the field theory we started with is in a non-compact space and the various symmetries are non-compact, one might prefer to perform this process in two steps. First, we follow the steps above to find the theory on a torus with periodic boundary conditions.  This is simpler than the more general case above and is generated by the (non-compact) group $\scK'=\bZ\otimes \bZ\subset K$ of pure translations. As in \eqref{eq:contsym}, its symmetry is the compact group $K'_\text{compact} = N_K(\scK')/\scK'$. Then, in the second step (which involves more complicated twists), we find the theory of interest by performing another quotient, this time by a finite subgroup of $K'_\text{compact}$.

\subsubsection{The inequivalent twisted models}
\label{sec:equivcont}

As we discussed, we use the pair $(k_1, k_2)$ to specify the boundary condition and, hence, the full model on the compact manifold. Some of the pairs lead to equivalent models, and we would like to identify the inequivalent ones.

First, if two pairs generate the same group $\scK$, they specify the same model.\footnote{The two sets of generators of the same $\scK$ are connected by the cycle redefinition $k'_I=k_1^{X^1_I}k_2^{X^2_I}$with fundamental domain shifted $\cs'^i_J = \cs^i_I X^I_J$. The allowed $X^I_J$ form the automorphism of $\scK$. For the torus, where $\scK \cong \pi_1(\Ts)$, one has $\Aut\left(\pi_1(\Ts)\right) \cong \mathrm{GL}(2,\bZ)$. For the Klein bottle, the fundamental group $\scK \cong \pi_1(\Kb)$ is not abelian, and hence the automorphism is the Borel subgroup of $\mathrm{GL}(2,\bZ)$,
\ie
\Aut\left(\pi_1(\Kb) \right) \cong
  \left\{\left.\left(
    \begin{matrix}
      a & b\\
      0 & c
    \end{matrix}
  \right)\right|a=\pm1,c=\pm 1, b \in \mathbb{Z}\right\}.
\fe}
In addition, we can also conjugate $k_1 $ and $k_2$ by an element $k \in K$. If this conjugation does not leave $\scK$  invariant, the model looks different, but it is related to the original one by a unitary transformation.

We end this subsection with an example. Consider a torus with a rectangular fundamental domain $\cl_1 \times \cl_2$. The twist is
\ie
\scK = \langle k_1,k_2\rangle\,,\quad k_1 = e^{ i\cl_1P_1}\,,\quad k_2 = e^{i\cl_2P_2}\,.
\fe
When $\cl_1=\cl_2$, the $\pi/2$ rotation exchanges the two cycles
\ie
e^{i\frac{\pi}{2}\sL} k_1 e^{-i\frac{\pi}{2}\sL}=k_2\,,\qquad e^{i\frac{\pi}{2}\sL} k_2 e^{-i\frac{\pi}{2}\sL}=k_1^{-1}\,.
\fe
and hence is a symmetry.

When $\cl_1 \neq \cl_2$,
\ie
e^{i\frac{\pi}{2}\sL} k_1 e^{-i\frac{\pi}{2}\sL}=e^{i \cl_1P_2}\,,\qquad e^{i\frac{\pi}{2}\sL} k_2 e^{-i\frac{\pi}{2}\sL}=e^{-i \cl_2P_1}\,,
\fe
and hence the $\frac{\pi}{2}$ rotation is not a symmetry.  Instead, it maps the model with $\scK$ to a unitarily equivalent model with a different $\scK$.

In what follows, we will analyze all models specified by $\scK$ up to conjugations.

\subsection{Parity anomaly in the five twisted models}
\label{sec:paritanomaly}

In this subsection, we will probe the parity anomaly \cite{Niemi:1983rq,Redlich:1983kn,Redlich:1983dv,Alvarez-Gaume:1984zst,Rao:1986ba,Witten:2015aba,Witten:2016cio} by deriving the projective representation of $K_\text{compact}$ under different twists. We will follow the analysis of \cite{Delmastro:2021xox}, who analyzed Majorana fermions on a torus and examined the projective phases in the algebra generated by $\sT$ and $(-1)^F$. See more details in Appendix \ref{app:twistedcontinuum}.

We will study five distinct cases of twists under which the symmetry operators are anomalous and are inequivalent under conjugation. To preserve the reflection symmetry and for the simplicity of the presentation, we set the shear parameter of the torus to zero.  Hence, we consider the identifications:
\ie \label{eq:bestshape}
& \Ts:\qquad (x^1,x^2)\sim (x^1+\cl_1,x^2)\sim (x^1,x^2+\cl_2)\,,\\
& \Kb:\qquad (x^1,x^2) \sim (x^1+\cl_1,x^2)\sim (-x^1,x^2+\cl_2)\,.
\fe
More general cases do not lead to new anomalies. See more discussions in Appendix \ref{app:decode}.

In the following list, we suppress the translation part of the generators that label the cases.  For example, the twists $k_1=e^{i \cl_1 P_1}$ and $k_2=e^{i \cl_2 P_2}$ are denoted as $(1,1)$; the twists $k_1=e^{i \cl_1 P_1}$ and $k_2=e^{i \cl_2 P_2}\Gamma$ are denoted as $(1,\Gamma)$; the twists $k_1=e^{i \cl_1 P_1}$ and $k_2=e^{i \cl_2 P_2}\sR$ are denoted as $(1,\sR)$.

\subsubsection{Torus}

\paragraph{Twists by $(1,1)$}
The symmetry generators are $e^{i\frac{2\pi}{\cl^i} \rho^i\bar P_i}$, $\Gamma$, $e^{i \theta \sJ}$, $\sR$, $\Xi$, and for $\cl_1=\cl_2$, also $e^{i \frac{\pi}{2} \sL}$. Here, we use a different normalization of the momentum operator $\bar P_i = \frac{\cl^i}{2\pi} P_i$ such that its eigenvalues are integers. (For the case with nonzero shear, see Appendix \ref{app:decode}.) The nontrivial projective relations among these generators are\footnote{Here, and in similar expressions below, the projective phases are determined by the quantum mechanics of the fermion zero modes.  See \cite{Seiberg:2025zqx} for more details.\label{detproph}}
\ie\label{eq:phasecont11}
    e^{i\pi \sJ}& = (-1)^{N_f}e^{2\pi i \sL}\,,\\
    \Gamma\sR &= (-1)^{N_f}\sR \Gamma\,,  \\
    \Xi \Gamma & = (-1)^{N_f} \Gamma \Xi  \,,\\
    \Xi \sR & = (-1)^{N_f} \sR \Xi\,,  \\
    \Xi^2 & = (-1)^{N_f}\,,
\fe
and hence the projective phases have order $2$.\footnote{As always, the operators can be re-phased, thus moving the projective phases around, but they cannot all be eliminated.\label{rephasing}} (Note that if we ignore $\Gamma$, $\sR$ and $\Xi$, the phase in the first equation can be redefined away, but this is not the case when the other symmetries are present.)

\paragraph{Twists by $(1,\Gamma)$}
Here, the symmetry generators are $e^{i \frac{2\pi}{\cl^i}\rho^i \bar P_i}$, $\Gamma$, $(-1)^F$, $\sR$, $e^{i \pi \sL}$, and $\Xi$. The projective relations read
\ie \label{eq:phasecont1g}
\sR (-1)^F &= (-1)^{N_f} (-1)^F \sR\,,\\
\Xi (-1)^F & = (-1)^{N_f} (-1)^F \Xi\,,\\
\Xi^2 &= (-1)^{\frac{N_f(N_f-1)}{2}}\,, \\
\Xi \sR & = (-1)^{\frac{N_f(N_f-1)}{2}} \sR\Xi\,,.\\
\fe
(See footnote \ref{detproph}.) Hence, the projective phases have order $4$.

The projective phases in these two cases of $(1,1)$ and $(1,\Gamma)$ are consistent with \cite{Delmastro:2021xox}, which computed the projective phase between $\sT$ and $(-1)^F$ for Majorana fermions on a torus.

\subsubsection{Klein bottle}\label{KBpa}

\paragraph{Twists by $(1,\sR)$}
Here, the symmetry generators are $e^{i\pi \bar P_1}$, $e^{i\frac{2\pi}{\cl_2}\rho \bar P_2}$, $\Gamma$, $e^{i \theta \sJ}$, $\sR$, and $\Xi$.\footnote{Because of the reflection twist along cycle 2, the continuous translation symmetry is broken to $\bZ_2$ generated by $e^{i\pi \bar P_1}$.  This is to be contrasted with the translation along cycle 2, $e^{i\epsilon\bar P_2}$ with $\epsilon \sim \epsilon+4\pi$.  Then $\epsilon=2\pi$ corresponds to $\sR$.} The nontrivial projective relations are\footnote{When $N_f$ is odd, we also have $\Gamma e^{i\theta \sJ}=e^{-i\theta}e^{-i\theta \sJ}\Gamma$ and $\Xi e^{i\theta \sJ}=e^{i\theta}e^{i\theta \sJ}\Xi$, where $e^{i\theta \sJ}$ is defined such that $e^{2\pi i\sJ}=1$. The expressions $\Gamma (-1)^F = -(-1)^F \Gamma$ and $\Xi (-1)^F = - (-1)^F \Xi$ are special cases of them. }
\ie \label{eq:phasecont1r}
\Gamma (-1)^F &= (-1)^{N_f} (-1)^F \Gamma\,,\\
\Xi (-1)^F & = (-1)^{N_f} (-1)^F \Xi\,,\\
\Xi^2 &= (-1)^{\frac{N_f(N_f-1)}{2}}\,, \\
\Xi \Gamma & = (-1)^{\frac{N_f(N_f-1)}{2}} \Gamma \Xi\,,\\
e^{2\pi i \bar P_2} \sR & = e^{-i \frac{2\pi}{8}N_f}\,,
\fe
from which we see that the projective phases are of order $4$. For more details of the projective phase $e^{i \frac{2\pi}{8}N_f}$ in the last equation, see Appendix \ref{app:kelinreal}.

\paragraph{Twists by $(\Gamma,\sR)$}
Here, the symmetry generators are $e^{i\frac{2\pi}{\cl_2}\rho \bar{P}_2}$, $\Gamma$, $(-1)^F$, $\sR$, and $\Xi$. When $N_f$ is even, the projective relations are
\ie \label{eq:phasecontgrA}
  \Xi (-1)^F & = (-1)^{\frac{N_f}{2}} (-1)^F \Xi\,,\\
  \Xi^2 & = (-1)^{\frac{N_f(N_f-2)}{8}}\,,
\fe
When $N_f$ is odd, $(-1)^F$ is not defined.\footnote{In this case, because of the Klein bottle geometry, the continuous rotation is no longer a symmetry. Thus, $(-1)^F$, an element in the spatial rotation, being undefined, does not raise any contradiction. However, as has been emphasized by various authors, including  \cite{Stanford:2019vob,Delmastro:2021xox,Witten:2023snr,Seiberg:2023cdc,Freed:2024apc,Harlow:2025cqc}, it might lead to more subtle problems.} For all $N_f$, we also have
\ie \label{eq:phasecontgrB}
e^{2\pi i \bar P_2} \sR & = e^{-i \frac{2\pi}{16}N_f}\,.
\fe
We conclude that the projective phases are of order $8$.

\paragraph{Twists by $(1,\Gamma\sR)$}
Here, the symmetry generators are $e^{i\pi \bar P_1}$, $e^{i\frac{2\pi}{\cl_2}\rho \bar{P}_2}$, $\Gamma$, $(-1)^F$, $\sR$, $e^{i\frac{\pi}{2}\sJ}e^{i \pi \sL}$, and $\Xi$. The corresponding projective relations are
\ie \label{eq:phasecont1gr}
\Gamma (-1)^F &= (-1)^{N_f} (-1)^F \Gamma\,, \quad
& \Xi \Gamma &= (-1)^{\frac{N_f(N_f-1)}{2}} \Gamma \Xi\,,\\
\sR (-1)^F &= (-1)^{N_f} (-1)^F \sR\,, \quad
& \Xi \sR &= (-1)^{\frac{N_f(N_f-1)}{2}} \sR \Xi\,,\\
\left(e^{i\frac{\pi}{2}\sJ}e^{i \pi \sL}\right) (-1)^F &= (-1)^{N_f} (-1)^F
\left(e^{i\frac{\pi}{2}\sJ}e^{i \pi \sL}\right)\,, \quad
& \Xi^2 &= (-1)^{\frac{N_f(N_f-1)}{2}}\,,\\
\left(e^{i\frac{\pi}{2}\sJ}e^{i \pi \sL}\right) \Xi &= (-1)^{\frac{N_f(N_f-1)}{2}}\Xi 
\left(e^{i\frac{\pi}{2}\sJ}e^{i \pi \sL}\right)\,, \quad
& \Xi (-1)^F &= (-1)^{N_f} (-1)^F \Xi\,.\\
\fe
which implies that the projective phases have order $4$.

\subsubsection{Conclusions}
\label{sec:paritanomalyconclusion}
This discussion (see, in particular \eqref{eq:phasecontgrA}) shows that the anomaly of the continuum Dirac fermion system is at least of order 8. 

What can this teach us about the problem of Majorana fermions?  Naively, our continuum theory involves two Majorana fermions, and one might be tempted to conclude that we derive a modulo 16 anomaly of the Majorana fermion problem.  This conclusion is incorrect since our modulo 8 anomaly involves internal symmetries that are not present in the single Majorana problem.  Indeed, even though the projective phases in \eqref{eq:phasecontgrA} do not involve such an internal symmetry, the twist by $\Gamma$ does.  Therefore, we can conclude that the order of the anomaly of the continuum Majorana fermion is at least 8.\footnote{One way to see that is to look at the equations that are independent of $\Gamma$ in \eqref{eq:phasecont1r}, whose phases vanish for $N_f=4$ and hence for $8$ Majorana fermions.}  Our method is limited to the compact flat spacetime in 2+1d and hence is not sensitive enough to detect the more subtle modulo 16 anomaly of Majorana fermions in \cite{Witten:2015aba,Witten:2016cio,Wang:2016qkb,Tachikawa:2016cha,Tachikawa:2016nmo}. 

Finally, we connect the above discussion about twists in space and projective representation of symmetry operators to the Euclidean spacetime interpretation. For Hilbert space twisted by $(k_1,k_2)$, we can consider the thermal partition function with insertion of symmetry operator $k_0$,
\ie
Z(k_1,k_2,k_0) = \mathrm{Tr}\left[e^{-\beta H(k_1,k_2)}k_0(-1)^F\right]\,.
\fe
If we insert $kk^{-1}=1$ with $k$ a symmetry operator and cycle $k$ around the trace, then the projective phases signal that the partition function vanishes. This partition function can also be interpreted as the partition function of the Dirac fermion defined on a 3d spacetime flat manifold twisted by $(k_1,k_2,k_0)$. Through modular transformations, this partition function can be interpreted with the temporal twist $k_1$ or $k_2$. In any interpretation, the number of spacetime zero modes remains the same.  And the partition function vanishes unless one explicitly inserts a Majorana field for each spacetime zero mode to saturate the integration measure.

\section{Lattice staggered fermions in 2+1d}
\label{sec:staggeredfermions}

\subsection{Staggered fermion, twists and symmetries}
\label{sec:presentation}
\subsubsection{Staggered fermion on the infinite lattice}
We start this section by reviewing the staggered lattice fermions system \cite{Kogut:1974ag,Banks:1975gq,Banks:1976ia,Susskind:1976jm,Affleck:2017ubr} and its symmetry \cite{Kluberg-Stern:1983lmr, Parisi:1983rv, Golterman:1984dn, Golterman:1985dz, Golterman:1986jf, Kilcup:1986dg}. This is a lattice model of a single Majorana fermion per site, coupled to a $\mathbb{Z}_2$ background gauge field with $\pi$-flux through each plaquette \cite{Affleck:2017ubr}.

We begin with an infinite square lattice, $\mathbb{Z}^2$, with $N_f$ Majorana fermions per site.\footnote{More precisely, the infinite lattice system should be defined as a limit of a finite system with some boundary conditions. Below, we will analyze various such boundary conditions.  For the purpose of the discussion in this subsection, we will ignore subtleties of the infinite volume system.} From here on, we'll omit the $N_f$ copies, and only write the $N_f$ dependence explicitly when it gives nontrivial results.

The only internal symmetry is the $\mathbb{Z}_2^F$ fermion parity,\footnote{Here we treat the lattice $(-1)^F$ symmetry as an internal symmetry.  This is consistent with the fact that the fermions are scalars on each site, i.e., they are invariant under site-centered rotations. We will address related subtleties when we match with the continuum in Section \ref{sec:tocontinuum}.} which acts as
\ie \label{eq:fermionparity}
    (-1)^F \psi_{\vec{\ell}} (-1)^F = -\psi_{\vec{\ell}}\,.
\fe
We turn on a background gauge field for this symmetry $\Upsilon_{\vec{\ell},\vec{\ell}+\hat{\mu}}$, with $\pi$-flux through each plaquette
\ie
&\Upsilon_{\vec{\ell},\vec{\ell}+\hat{1}}\Upsilon_{\vec{\ell}+\hat{1},\vec{\ell}+\hat{1}+\hat{2}}\Upsilon_{\vec{\ell}+\hat{1}+\hat{2},\vec{\ell}+\hat{2}}\Upsilon_{\vec{\ell}+\hat{2},\vec{\ell}}=-1\\
&\Upsilon_{\vec{\ell},\vec{\ell}+\hat{\mu}}\in \mathbb{Z}_2\,,
\fe
and study the Hamiltonian 
\ie \label{eq:staggeredH}
  H = i \sum_{A=1}^{N_f}\sum_{\vec{\ell} \in \mathbb{Z}^2} \sum_{\mu=1,2}
\Upsilon_{\vec{\ell},\vec{\ell}+\hat{\mu}}\, \psi_{\vec{\ell}}^A\psi_{\vec{\ell}+\hat{\mu}}^A \,.
\fe
The gauge transformation of the background field acts as
\ie \label{eq:gaugetransform}
    &\psi_{\vec{\ell}}^A \to \lambda(\vec{\ell}) \psi_{\vec{\ell}}^A\,, \\
&\Upsilon_{\vec{\ell},\vec{\ell}+\hat{\mu}} \to
   \lambda(\vec{\ell}) \lambda(\vec{\ell}+\hat{\mu}) \Upsilon_{\vec{\ell},\vec{\ell}+\hat{\mu}}\,,\\
&\lambda(\vec{\ell}) =\pm 1\,.
\fe
We emphasize that since this transformation involves an action on the background fields $\Upsilon_{\vec{\ell},\vec{\ell}+\hat{\mu}} $, it is not a symmetry of the problem. Equivalently, we can think of the classical gauge field $\Upsilon_{\vec{\ell},\vec{\ell}+\hat{\mu}}$ as a spurion for this gauge symmetry.

Using this gauge freedom, we choose 
\ie\label{UIpsilonchoice}
\Upsilon_{\vec{\ell},\vec{\ell}+\hat{\mu}}= \eta_\mu(\vec{\ell})
   = (-1)^{\sum_{\nu < \mu}\ell^\nu} =\begin{cases} 1&\mu=1\\ (-1)^{\ell_1}&\mu=2 \end{cases} \,.
\fe
Then the crystalline symmetries consist of lattice translations $T_{1,2}$, a site-centered reflection $\R$ (along the 1-axis), and a site-centered rotation $\C$, acting as
\ie \label{eq:cryssym1}
    \begin{aligned}
      T_1 \psi_{\vec{\ell}} T_1^{-1} &= (-1)^{\ell^2} \psi_{\vec{\ell}+\hat{1}}\,, &\quad
      T_2 \psi_{\vec{\ell}} T_2^{-1} &= \psi_{\vec{\ell}+\hat{2}}\,, \\
      \R \psi_{\vec{\ell}} \R^{-1} &= (-1)^{\ell^1} \psi_{-\ell^1, \ell^2}\,, &\quad
      \C \psi_{\vec{\ell}} \C^{-1} &= (-1)^{\ell^1 \ell^2+\ell^2} \psi_{-\ell^2, \ell^1}\,.
    \end{aligned}
\fe
In addition, there is an anti-unitary time-reversal symmetry $\T$,
\ie \label{eq:cryssym2}
    \T \psi_{\vec{\ell}} \T^{-1}=(-1)^{\ell^1+\ell^2}\psi_{\vec{\ell}}\,.
\fe

The non-trivial algebraic relations of these symmetries are \cite{Kluberg-Stern:1983lmr,Parisi:1983rv,Golterman:1984dn,Golterman:1985dz,Golterman:1986jf,Kilcup:1986dg},
\ie \label{eq:algebrauntwisted}
    \begin{gathered}
        \C^4=1\,,\quad \C T_1 \C^{-1}=T_2\,,\quad \C T_2 \C^{-1}=(-1)^F T_1^{-1}\,,\quad T_1 T_2=(-1)^F T_2 T_1\,,\\
        \R^2=1\,,\quad T_1 \R=(-1)^F \R T_1^{-1}\,,\quad \R \C \R^{-1}=\C^{-1}\,,\\
        \T^2=1\,,\quad \T T_i =(-1)^F T_i \T\, .
    \end{gathered}
\fe
Note that both $\T$ and $\R$ may be redefined by multiplication with $(-1)^F$ without altering the algebra.

The factors of $(-1)^F$ in the group algebra reflect the presence of $\pi$-flux in the background $\mathbb{Z}_2$ gauge field. Since the site-centered rotation satisfies $\C^4=1$, the Majorana fermion at each lattice site transforms as a scalar. It is often useful to consider instead the link-centered reflection $\R'=T_1 \R$ and the plaquette-centered rotation $\C'=T_1 \C$, for which the algebra becomes
\ie
    (\R')^2=(-1)^F,\qquad (\C')^4 = (-1)^F\, .
\fe

More generally, for any background gauge field $\tilde \Upsilon_{\vec{\ell},\vec{\ell}+\hat{\mu}}$ related by the gauge transformation \eqref{eq:gaugetransform}, the symmetry actions are conjugated by the operator that implements $\psi_{\vec{\ell}}\to \lambda(\vec{\ell}) \psi_{\vec{\ell}}$. This does not change the algebra.

The translation operators $T_{1,2}$ generate a central extension $\Lambda$ of $\mathbb{Z}^2$ by $\mathbb{Z}_2^F$. The full symmetry group of the infinite lattice is 
\ie \label{eq:G}
  G = \Lambda \rtimes \left( D_4 \times \mathbb{Z}_2^\Omega \right) \,,
\fe
where $D_4$ is the dihedral group of the square, generated by $\C$ and $\R$, and $\mathbb{Z}_2^\Omega$ is generated by $\Omega=\T\C^2$. Altogether, we have the exact sequence
\ie
  1 \;\longrightarrow\; \mathbb{Z}_2^F
   \;\longrightarrow\; G
   \;\longrightarrow\; \mathbb{Z}^2 \rtimes \left( D_4  \times \mathbb{Z}_2^\Omega \right)\;\longrightarrow\; 1 \,,
\fe
where $\mathbb{Z}^2 \rtimes \left( D_4  \times \mathbb{Z}_2^\Omega \right)$ is the crystalline symmetry group of the two-dimensional infinite lattice in the absence of fermionic degrees of freedom (as well as the background gauge fields).

\subsubsection{Quotient to finite lattice}
\label{sec:finitelattice}
Mirroring the continuum discussion in Subsection \ref{sec:conttwist}, here, we repeat it on the lattice.  We identify sites on the infinite lattice to obtain compact flat lattice models. 

This identification can be made by choosing an appropriate subgroup $\scG$ of the symmetry of the problem $G$ in \eqref{eq:cryssym1}.  $\scG$ is generated by two elements $g_I\in G$ with $I=1,2$.  Then, we impose
\ie
\psi_{\vec{\ell}} = g_I \psi_{\vec{\ell}} g_I^{-1} \,.
\fe
The fact that $g_I$ are symmetry operators guarantees that the local Hamiltonian on each link is properly identified. We choose the subgroup $\scG = \langle g_1, g_2 \rangle$ to be isomorphic to the fundamental group $\scG \cong \pi_1(\Ts)$ or $\scG \cong \pi_1(\Kb)$, such that the finite lattice $\bZ^2/\scG$ is a torus or a Klein bottle.

Geometrically, for each $g_I$, we identify the sites $\vec \ell \sim \vec\ell'$ and the links $\vec \ell +\hat \mu \sim \vec \ell'+\hat \mu'$.
(Note that $\hat \mu$ can differ from $\hat \mu'$, as the identification can involve reflection along the sides of the lattice or along diagonals.)  This leads to
\ie \label{eq:psitrans}
&\psi_{\vec{\ell}} = \lambda_I(\vec{\ell}') \psi_{\vec{\ell}'}\,. \\
\fe
and 
\ie  \label{eq:alphaperiodic}
&\Upsilon_{\vec{\ell},\vec{\ell}+\hat{\mu}}\psi_{\vec{\ell}}\psi_{\vec{\ell}+\hat{\mu}}=\lambda_I(\vec{\ell}')\lambda_I(\vec{\ell}'+\hat{\mu}')\Upsilon_{\vec{\ell}',\vec{\ell}'+\hat{\mu}'}\psi_{\vec{\ell}'}\psi_{\vec{\ell}'+\hat{\mu}'}\,,
\fe

On the torus, this map does not involve orientation reversal, then \eqref{eq:alphaperiodic} reduces to
\ie \label{eq:samefield}
\Upsilon_{\vec{\ell},\vec{\ell}+\hat{\mu}}=\lambda_I(\vec{\ell}')\lambda_I(\vec{\ell}'+\hat{\mu}')\Upsilon_{\vec{\ell}',\vec{\ell}'+\hat{\mu}'}\,.
\fe

However, on the Klein bottle, the direction of certain links can be reversed. Consequently, the background gauge fields given by \eqref{eq:alphaperiodic} acquire a sign change
\ie \label{eq:oppfield}
\Upsilon_{\vec{\ell},\vec{\ell}+\hat{\mu}}=-\lambda_I(\vec{\ell}')\lambda_I(\vec{\ell}'+\hat{\mu}')\Upsilon_{\vec{\ell}',\vec{\ell}'+\hat{\mu}'}\,.
\fe
(A similar issue arose in Appendix E.2 in \cite{Seiberg:2025zqx}.)

The relations \eqref{eq:samefield} and \eqref{eq:oppfield} can be interpreted as transition functions for the background $\Upsilon$.  Interestingly, it is not a standard $\bZ_2$ lattice gauge field, since $\Upsilon$ can change sign under reflection. Relatedly, depending on how we define the local flux around some plaquettes, it is no longer $\pi$, but it vanishes.  We see that the cycle involving a reflection can be thought of as leading to a ``reflection defect'' along which we have plaquettes with vanishing flux (as opposed to $\pi$ flux).\footnote{One might be concerned that this defect, which is associated with a strip of such plaquettes, is incompatible with the translation symmetry along this cycle.  However, as we have checked, our infinite lattice system is translation-invariant, and so are the boundary conditions along this cycle.  Consequently, the location of the ``reflection defect'' can be changed by a unitary transformation.}

We can now apply the group-theoretic arguments in Subsection \ref{sec:conttwist}. The remaining symmetry on the finite lattice is
\ie \label{eq:latcompsym}
G_\text{compact} = {N_G(\scG)\over \scG }\,.
\fe
As in Subsection \ref{sec:equivcont}, redefinitions of the cycles $(g_1,g_2)$ and conjugation by elements $g \notin N_G(\scG)$ yield equivalences between different models.

As in the comment at the end of Section \ref{symmetryoftwistedcon}, we can perform this discussion in two steps.  First, we derive the simple theory on a square torus with an even number of sites along each direction and with periodic boundary conditions, whose symmetry group is finite.  And then, we perform another quotient by a finite group to find the model of interest.

\subsubsection{Lattice on torus and Klein bottle}
\label{sec:latTsandKb}

As derived in Appendix \ref{app:fundgroup}, we present an overcomplete list of models and organize the generators such that the resulting fundamental domains mirror the continuum tori and Klein bottles in \eqref{simplT2K2}. This partially identifies equivalent models, but we do not restrict the list to a single representative in each equivalence class.

For the torus, we can take the generators to be
\ie \label{eq:T}
g_1 = T_1^{L_1}\left[(-1)^F\right]^{W_1}\,,\quad
g_2 = T_1^{b}T_2^{L_2}\left[(-1)^F\right]^{W_2}\,,
\fe
where $L_I$ and $b$ are integers, such that  $0\le b<L_1$, $0 < L_2 $, and $W_I=0,1$.  The product $L_1L_2$ has to be even to ensure that $g_1g_2=g_2g_1$.\footnote{\label{ft:shape1} Equivalently, this relation follows from the consistency of the background $\Upsilon_{\vec{\ell},\vec{\ell}+\hat{\mu}}$ on the finite lattice.  To see that, we calculate the total flux through the torus in two ways.  First, we multiply the contributions from all the plaquettes to find $(-1)^{L_1L_2}$.  Second, we note that every link appears twice; hence, the total flux must equal $1$.}  Interestingly, this conclusion guarantees that the total number of fermions in our finite lattice is even, thus avoiding the issues discussed in \cite{Stanford:2019vob,Delmastro:2021xox,Witten:2023snr,Seiberg:2023cdc,Freed:2024apc,Seiberg:2025zqx,Harlow:2025cqc}.

For the Klein bottle, we can take the generators to be ($L_I>0$)
\ie \label{eq:K1}
g_1 = T_1^{L_1}\left[(-1)^F\right]^{W_1}\,, \quad
g_2 = T_2^{L_2}\,\R\left[(-1)^F\right]^{W_2}\,,
\fe
or
\ie \label{eq:K2}
g_1 = T_1^{L_1}\left[(-1)^F\right]^{W_1}\,, \quad
g_2 = T_2^{L_2}(T_1\R)\left[(-1)^F\right]^{W_2}\,,
\fe
with the constraint that $L_1(L_2-1)$ is even to ensure that $g_2g_1g_2^{-1}=g_1^{-1}$.\footnote{\label{ft:shape2} Equivalently, as in footnote \ref{ft:shape1}, it follows from computing the total flux through the space on the finite lattice.  Unlike the case of the torus, here, we have $L_1$ edges that are glued using reflection, and hence their $\Upsilon$ has an additional minus sign that changes the sign of flux. (See discussion around \eqref{eq:oppfield}).  This leads to $(-1)^{L_1(L_2-1)}=1$, and hence, $L_1(L_2-1)$ should be even.}   Note that if $L_1$ is even, then the total number of fermions is even and we avoid the issues in \cite{Stanford:2019vob,Delmastro:2021xox,Witten:2023snr,Seiberg:2023cdc,Freed:2024apc,Seiberg:2025zqx,Harlow:2025cqc}.  However, when $L_1$ is odd, this is not the case.  We will return to this case in Subsection \ref{sec:latticeanomaly}.

In addition, we can take the generators ($L_I>0$)
\ie \label{eq:K3}
g_1 = T_1^{L_1} T_2^{L_1}\left[(-1)^F\right]^{W_1}\,, \quad
g_2 = T_1^{L_2} T_2^{-L_2} (\C\R)\left[(-1)^F\right]^{W_2}\,,
\fe
with no restriction on $L_1$ or $L_2$; or
\ie \label{eq:K4}
g_1 = T_1^{L_1} T_2^{L_1} \left[(-1)^F\right]^{W_1}\,, \quad
g_2 = T_1^{L_2} T_2^{-L_2} (T_2\C\R)\left[(-1)^F\right]^{W_2}\,,
\fe
with the condition that $L_1$ is even to ensure that $g_2g_1g_2^{-1}=g_1^{-1}$.\footnote{\label{ft:shape3}  Equivalently, we can follow the analysis in footnotes \ref{ft:shape1} and \ref{ft:shape2}.  In \eqref{eq:K3} and \eqref{eq:K4}, the geometry is more complicated, but again, the total number of plaquettes plus the number of flipped edges should be even.}

In \eqref{eq:K1} - \eqref{eq:K4}, we did not include the reflection twist  $\C^2\R$ or $\C^3\R$, because they are equivalent to the ones in \eqref{eq:K1} - \eqref{eq:K4} using conjugation with $\C$.  More details are provided in Appendix \ref{app:decode}.

\subsection{The anomaly}
\label{sec:latticeanomaly}
Here, we find the anomaly of the staggered fermion with various twists. We do that by identifying the projective representation of the symmetry operators. The nonzero fermion modes cannot contribute to the projective phases; therefore, we can focus solely on zero modes.\footnote{This point reflects a general principle about anomalies. They are the same at high energies and at low energies.  Focusing on the zero modes amounts to taking the extreme low-energy limit.} (See more details in Appendix \ref{app:twistedcontinuum}).

Next, we list seven cases where the fermion zero modes make the symmetry operators act projectively. For this purpose, we identify cases that differ by conjugation and have the same values of $S^i_I\mod 2$. (See more details in Appendix \ref{app:twistedcontinuum}.)

For each case, we pick a pair of representative generators $g_{1,2}$ and write the power of $T_{1,2}$ modulo 2.  We also choose the shape of the fundamental domain that preserves as many symmetries as possible. This leads us to the following cases:
\begin{itemize}
    \item for the torus, we choose (as in \eqref{eq:T})
    \ie \label{eq:shapeT}
    \psi_{\ell^1,\ell^2} = (-1)^{\ell^2L_1}\psi_{\ell^1+L_1,\ell^2} = \psi_{\ell^1,\ell^2+L_2}\,.
    \fe
    \item for the Klein bottle
    \begin{itemize}
    \item with reflection twist $\R$, we choose (as in \eqref{eq:K1})
    \ie \label{eq:shapeK1a}
    \psi_{\ell^1,\ell^2} = (-1)^{\ell^2L_1}\psi_{\ell^1+L_1,\ell^2} = (-1)^{\ell^1}\psi_{-\ell^1,\ell^2+L_2}\,,
    \fe
    or (as in \eqref{eq:K2})
    \ie \label{eq:shapeK1b}
    \psi_{\ell^1,\ell^2} = (-1)^{\ell^2L_1}\psi_{\ell^1+L_1,\ell^2} = (-1)^{\ell^1+\ell^2}\psi_{-\ell^1+1,\ell^2+L_2}\,.
    \fe
    \item with reflection twist $\C\R$, we  (as in \eqref{eq:K3})\footnote{One can also consider $  \psi_{\ell^1,\ell^2} = (-1)^{\ell^2L_1}\psi_{\ell^1+L_1,\ell^2+L_1} = (-1)^{\ell^1\ell^2+\ell^1(L_2+1)+\ell^2}\psi_{-\ell^2+L_2,-\ell^1-L_2-1}\,$. However, such boundary conditions do not lead to zero modes, and therefore, there are no projective phases.}
    \ie \label{eq:shapeK2}
    \psi_{\ell^1,\ell^2} &= (-1)^{(\ell^2+L_1)L_1}\psi_{\ell^1+L_1,\ell^2+L_1} \\
    & = (-1)^{\ell^1\ell^2+(\ell^1+L_2)L_2+\ell^2+\ell^1}\psi_{-\ell^2+L_2,-\ell^1-L_2}\,.
    \fe
    Geometrically, these two types of Klein bottle differ by $\frac{\pi}{4}$ rotation, which is not part of the crystalline symmetries, and therefore these cases are distinct.
\end{itemize}
\end{itemize}

\subsubsection{Torus}

Here, we use the shape specified in \eqref{eq:shapeT}.  As we discussed in Footnote \ref{ft:shape1}, here, we need $L_1L_2$ to be even.

\paragraph{No Twist $(1,1)$} By this, we mean that $L_1$ and $L_2$ are even.  In order to have maximal symmetry, we take $L_1=L_2$.  The remaining symmetries are generated by $T_{1}$, $T_2\R$, $\C$, $(-1)^F$, and $\Omega$. The nontrivial projective relations among these generators are
\ie \label{eq:phaselat11}
\C^4&=(-1)^{N_f}\,,\\
T_1(T_2\R) & = (-1)^{N_f}(T_2\R)T_1^{-1}\,,\\
\Omega T_1 & = (-1)^{N_f} T_1^{-1} \Omega\,,\\
\Omega (T_2\R) & = (-1)^{N_f}(\C T_1\C^{-1})^{-2}(T_2\R) \Omega \,,\\
\Omega^2 & = (-1)^{N_f}\,.
\fe
They are of order 2.

\paragraph{Twists by $(1, T_2)$} By this, we mean that $L_1$ is even and $L_2$ is odd. The remaining symmetries are generated by $T_1^2$, $T_2$, $T_2\R$, $T_1\C^2\R $, $(-1)^F$, and $\Omega$. The nontrivial projective phases are
\ie \label{eq:phaselat1g}
(T_2\R) (-1)^F & = (-1)^{N_f} (-1)^F (T_2\R)\,,\\
(T_1\C^2\R ) (-1)^F & = (-1)^{N_f} (-1)^F (T_1\C^2\R )\,,\\
\Omega^2 &  = (-1)^{\frac{N_f(N_f-1)}{2}}\,,\\
\Omega (T_2\R)& = (-1)^{\frac{N_f(N_f-1)}{2}} T_2^{-2}(T_2\R)\Omega\,,\\
\Omega (T_1\C^2\R ) &= (-1)^{\frac{N_f(N_f-1)}{2}} T_1^{-2}(T_1\C^2\R )\Omega\,.
\fe
These phases are of order 4.\footnote{As a check, we can use the fact that $L_2$ is odd and reduce this problem to the 1+1d Majorana chain with an even number of sites.  Then, these projective phases agree with \cite{Seiberg:2023cdc,Seiberg:2025zqx}.\label{reducetochane}}

\subsubsection{Klein bottle}
\bigskip
\centerline{\textit{Twists by $\R$}}
\bigskip
Here, we use \eqref{eq:shapeK1a} and \eqref{eq:shapeK1b}. As we discussed in Footnote \ref{ft:shape2}, we need $L_1(L_2-1)$ to be even.

\paragraph{Twists by $(1, T_2\R)$} By that, we mean that $L_1$ is even and $L_2$ is odd with \eqref{eq:shapeK1a}. The remaining symmetries are generated by $T_1^{\frac{L_1}{2}}$, $T_2$, $T_2\R$, $(-1)^F$, and $\Omega$. The nontrivial projective relations are
\ie \label{eq:phaselat1r}
T_1^{\frac{L_1}{2}}(-1)^F
&= (-1)^{\frac{N_fL_1}{2}}(-1)^F T_1^{\frac{L_1}{2}}\,, &
T_2(-1)^F
&= (-1)^{N_f}(-1)^F T_2\,,\\
\Omega (-1)^F
&= (-1)^{N_f}(-1)^F\Omega\,, &
\Omega^2
&= (-1)^{\frac{N_f(N_f-1)}{2}}\,,\\
\Omega T_1^{\frac{L_1}{2}}
&= (-1)^{\frac{N_f(N_f-1)}{2}\frac{L_1}{2}}T_1^{\frac{L_1}{2}}\Omega\,, &
\Omega T_2
&= (-1)^{\frac{N_f(N_f-1)}{2}}T_2^{-1}\Omega\,.
\fe
And the phases are of order 4.

\paragraph{Twists by $(T_1,T_2\R)$} By that, we mean that $L_1$ and $L_2$ are odd with \eqref{eq:shapeK1a}.  The remaining symmetries are generated by $T_2^2$, $T_2\R$, $(-1)^F$, and $\Omega$. This is the only case where, for odd $N_f$, the total number of fermions is odd.  (Compare with \cite{Seiberg:2023cdc,Seiberg:2025zqx}.) For even $N_f$, we have 
\ie \label{eq:phaselatgrA}
\Omega (-1)^F & = (-1)^{\frac{N_f}{2}}(-1)^F\Omega\,, \\
\Omega^2 & = (-1)^{\frac{N_f(N_f-2)}{8}}\,.
\fe
When $N_f$ is odd, $(-1)^F$ is not well-defined. These phases are of order 8.  

For odd $N_f$, we also have
\ie \label{eq:phaselatgrB}
T_2^{L_2}\R = e^{-i \frac{2\pi}{16}n}\,,
\fe
where $n$ is any odd number.  This phase is derived by calculating the vacuum momentum.\footnote{Alternatively, as in footnote \ref{reducetochane}, we can reduce our problem to the 1+1d Majorana chain with an odd number of sites, and use the results in \cite{Seiberg:2023cdc,Seiberg:2025zqx}. Other phases in \cite{Seiberg:2023cdc,Seiberg:2025zqx} are also consistent with \eqref{eq:phaselatgrA}.}  An analogous continuum calculation will be presented in Appendix \ref{app:kelinreal}. To obtain this specific phase $e^{-2\pi i n/16 }$, we adopt the phase redefinition of the translation operator detailed in \cite{Seiberg:2023cdc}, which shows that this projective phase is of order 2. 

\paragraph{Twists by $(1, \R )$} By that, we mean that $L_1$ and $L_2$ are even with \eqref{eq:shapeK1a}. The remaining symmetries are generated by $T_1^{\frac{L_1}{2}}$ (if $\frac{L_1}{2}$ is even), $T_2$, $T_2\R$, $\C^2$, $(-1)^F$, and $\Omega$. The nontrivial projective relations are
\ie \label{eq:phaselat1gr1}
T_2 (-1)^F &= (-1)^{N_f} (-1)^F T_2\,, &  \Omega T_2 & = (-1)^{\frac{N_f(N_f-1)}{2}} T_2^{-1} \Omega\,,  \\
(T_2 \mathcal{R}) (-1)^F &= (-1)^{N_f} (-1)^F (T_2 \mathcal{R})\,, & \Omega (T_2 \mathcal{R}) & = (-1)^{\frac{N_f(N_f-1)}{2}} T_2^{-2} (T_2 \mathcal{R}) \Omega\,, \\
\mathcal{C}^2 (-1)^F &= (-1)^{N_f} (-1)^F \mathcal{C}^2\,, & \Omega \mathcal{C}^2& = (-1)^{\frac{N_f(N_f-1)}{2}} \mathcal{C}^2 \Omega\,, \\
\Omega^2 &= (-1)^{\frac{N_f(N_f-1)}{2}}\,, & \Omega (-1)^F &  = (-1)^{N_f}(-1)^F \Omega\,.
\fe
These phases are of order 4.

\paragraph{Twists by $(1, T_1 T_2\R)$} By that, we mean that  $L_1$ is even and $L_2$ is odd with \eqref{eq:shapeK1b} where there is another $T_1$ shift in the second cycle. The remaining symmetries are generated by $T_1^{\frac{L_1}{2}}$, $T_2^2$, $T_1T_2\R$, $T_1\C^2$, $(-1)^F$, and $T_1\Omega$. The nontrivial projective relations are
\ie \label{eq:phaselat1gr2}
T_1^{\frac{L_1}{2}} (-1)^F & = (-1)^{\frac{N_fL_1}{2}} (-1)^FT_1^{\frac{L_1}{2}}\,, \\
T_1^{\frac{L_1}{2}}(T_1\Omega) & = (-1)^{\frac{N_f(N_f-1)}{2}\frac{L_1}{2}} (T_1\Omega) T_1^{-\frac{L_1}{2}}\,.
\fe
For $L_1=0\bmod 4$, there are no projective phases, and for $L_1=2\bmod 4$, the phases are of order $4$.  (Recall that $L_1$ is even.) 

 \bigskip\bigskip
\centerline{\textit{Twists by $\C\R$}}
\bigskip

Here, we use \eqref{eq:shapeK2}. As we discussed in Footnote \ref{ft:shape3}, there is no restriction on $L_1$ and $L_2$.

\paragraph{Twists by $(1,\C \R)$} By that, we mean that  $L_1$ and $L_2$ are even. The remaining symmetries are generated by $T_1^{\frac{L_1}{2}}T_2^{\frac{L_1}{2}}$ (if $\frac{L_1}{2}$ is even), $T_1^2T_2^{-2}$, $\C \R$, $\C^2$, $(-1)^F$, and $\Omega$. The projective relations are
\ie \label{eq:phaselat1gr3}
\C^2 (-1)^F & = (-1)^{N_f} (-1)^F\C^2\,,\\
\Omega \mathcal{C}^2& = (-1)^{\frac{N_f(N_f-1)}{2}} \mathcal{C}^2 \Omega\,, \\
\Omega^2 &= (-1)^{\frac{N_f(N_f-1)}{2}}\,, \\
\Omega (-1)^F &  = (-1)^{N_f}(-1)^F \Omega\,.
\fe
The phases are of order 4.

\subsubsection{Conclusions}

As we have seen, various spatial twists result in distinct projective phases.  The strongest constraint comes from the Klein bottle with the twist $(T_1,T_2\R)$ in \eqref{eq:phaselatgrA} and \eqref{eq:phaselatgrB}.  It determines that the anomaly is at least of order 8. This anomaly is also at most of order 8 because with $N_f = 0 \bmod 8$ one can gap all the degrees of freedom by onsite generic four fermion coupling that preserves all the crystalline symmetries, as discussed in \cite{Fidkowski:2009dba,Seiberg:2025zqx}.  We conclude that here, the anomaly is of order $8$.

\section{Matching the lattice and the continuum}
\label{sec:tocontinuum}
In this Section, we will match the lattice and the continuum results.  We will find the map between the twisted boundary conditions and the symmetry operators of these two problems.  Then, this map will enable us to verify that the anomaly in the lattice problem is accurately represented in the continuum.

In the continuum limit, the lattice spacing $a$ is taken to zero, and we focus on the low-lying modes to find a continuum field theory.  On a compact lattice, the $a \to 0$ is taken with fixed physical size $\cs^i_I = aS^i_I$.  While taking this limit, if $S^i_I\neq 0$, we take it to infinity without changing $S^i_I \bmod 2$.

\subsection{Untwisted staggered fermion}
\label{sec:untwistedcontinuum}

We start with untwisted staggered fermions on a torus, i.e., Majorana fermions on a lattice with periodic boundary conditions
\ie \label{eq:latticeshear}
\psi_{\vec{\ell}} = \psi_{\vec{\ell}+\vec{S}_1} = \psi_{\vec{\ell}+\vec{S}_2}\,,
\fe
with even $S_I^i$.

This untwisted model is solved in Appendix \ref{app:twistedcontinuum}. The low-energy modes fall into four classes, with momenta of the Fourier modes around the four corners of the Brillouin zone. We can relate the lattice momentum $\bar{P}_I$ to the continuum momentum $P_i$ by $\bar{P}_I = \frac{a}{2\pi}S^i_I P_i$. This leads to a single continuum of Dirac fermions. 

In Appendix \ref{app:twistedcontinuum}, we also map the actions of the lattice symmetry operators to those of the continuum theory:
\begin{itemize}
\item The lattice internal symmetry $(-1)^F$ is the same as the continuum fermion parity $(-1)^F$.
\item The lattice translation symmetries lead, in the continuum, to emanant internal symmetries \cite{Cheng:2022sgb,Seiberg:2023cdc,Seiberg:2025zqx}, $\Gamma_{1,2}$,
\ie \label{eq:transmap}
  T_1 \to e^{2\pi i (S^{-1})^I_1\bar P_I}\Gamma_1\quad, \quad
  T_2 \to e^{2\pi i (S^{-1})^I_2\bar P_I} \Gamma_2\,,
\fe
where to compare with \eqref{eq:O2}, use 
\ie
\Gamma_1=\Gamma e^{i\frac{\pi}{2}\sJ} \,, \quad  \Gamma_2 = \Gamma\,.
\fe
They obey the algebra
\ie\label{Gammaalgebra}
\Gamma_1^2=\Gamma_2^2=1,\quad \Gamma_1 \Gamma_2=(-1)^F \Gamma_2 \Gamma_1\,,
\fe
i.e., they form $D_4 \subset O(2)$.\footnote{In \cite{vandenDoel:1983mf,Kluberg-Stern:1983lmr, Parisi:1983rv, Golterman:1984dn, Golterman:1985dz,  Golterman:1986jf, Kilcup:1986dg}, similar factorization on the zero modes was also discussed in $3+1d$ complex staggered fermion.}  As a check, the lattice relations
\ie
T_1^{S^1_1}T_2^{S^2_1} = 1\,,\quad T_1^{S^1_2}T_2^{S^2_2} = 1\,,\quad T_1T_2 = (-1)^F T_2T_1
\fe
are consistent with the continuum relations \eqref{Gammaalgebra}. 
\item The lattice reflection and the spacetime-reversal operators are mapped as
\ie\label{eq:refomegamap}
    \R \to \Gamma_2\sR\qquad, \qquad
    \Omega \to \Xi\,.
\fe
\end{itemize}

The continuum fields can be redefined by conjugation by continuum transformation $k \in K$, and therefore the map from the lattice operators to the continuum operators is up to the conjugation by $k$. However, the continuum relations \eqref{Gammaalgebra} are unchanged.

Clearly, all the lattice symmetries are exact symmetries of the model, and the other continuum symmetries are approximate.  In particular, $\Xi$ is exact.  

However, some of the new emanant continuum symmetries are special, because they are not violated by any local operator in the continuum theory \cite{Cheng:2022sgb}. Examples are   $\Gamma_{1,2}$.  Since $\R$ is exact and $\Gamma_2$ is not violated by local operators, the same is true for $\sR$.  Indeed, $O(1/S^i_I)$ corrections would not be compatible with $\R^2=1$ for arbitrary $S_I^i$.

Finally, the lattice rotation symmetry leads to emanant continuum spatial and internal rotations
\ie \label{eq:rotmap}
    \C \to e^{i\frac{\pi}{4}\sJ}e^{i\frac{\pi}{2}\sL}\,.
\fe
$\C$ is exact, but unlike the other emanant symmetries, the separate factors $e^{i\frac{\pi}{4}\sJ}$ and $e^{i\frac{\pi}{2}\sL}$ can be violated by higher-dimensional operators.

To summarize, in the continuum limit, we have the following map
\ie\label{latticetocom}
G & \,\,\,\,\quad \longrightarrow && \qquad K \\
T_{1,2} &\qquad  \mapsto && \qquad\Gamma_{1,2}\,,\\
\R &\qquad  \mapsto && \qquad\Gamma_2\sR\,,\\
\C &\qquad  \mapsto && \qquad e^{i\frac{\pi}{4}\sJ}e^{i\frac{\pi}{2}\sL}\,,\\
\Omega & \qquad  \mapsto && \qquad\Xi\,.
\fe

\subsection{Mapping the twisted problems and matching the projective phases}
\label{sec:latconttwistmatch}

\begin{table}[t]
    \centering
  \begin{tabular}{|c|c|c|c|}
    \hline
    \multicolumn{3}{|c|}{Lattice} & \multirow{2}{*}{\makecell{Continuum\\Twist}} \\
    \cline{1-3}
    Boundary Condition & $L_1$  & $L_2$  & \\
    \hline
    \multirow{2}{*}{$(\ell^1,\ell^2) \sim (\ell^1+L_1,\ell^2) \sim (\ell^1,\ell^2+L_2)$}
        & even & even & $\left(1,1\right)$ \\
    \cline{2-4}
        & even & odd  & $\left(1,\Gamma\right)$ \\
    \hline
    \multirow{3}{*}{$(\ell^1,\ell^2) \sim (\ell^1+L_1,\ell^2) \sim (-\ell^1,\ell^2+L_2)$}
        & even & odd & $\left(1, \sR\right)$ \\
    \cline{2-4}
        & odd & odd  & $\left(\Gamma, \sR\right)$ \\
    \cline{2-4}
        & even & even & \multirow{3}{*}{$\left(1, \Gamma\sR \right)$} \\
    \cline{1-3}
    $(\ell^1,\ell^2) \sim (\ell^1+L_1,\ell^2) \sim (-\ell^1+1,\ell^2+L_2)$
        & even & odd & \\
    \cline{1-3}
    \multirow{1}{*}{$(\ell^1,\ell^2)\sim (\ell^1+L_1,\ell^2+L_1) \sim (-\ell^2+L_2,-\ell^1-L_2)$}
        & even & even & \\
    \hline
\end{tabular}
    \caption{The map between the twists on the lattice and in the continuum for the models with zero modes. Here we omit the phases when identifying fermions, which are specified in \eqref{eq:shapeT}, \eqref{eq:shapeK1a}, \eqref{eq:shapeK1b}, and \eqref{eq:shapeK2} (the rows in the first column are in the same order as there).}
    \label{tab:twistmatch}
\end{table}

Here, we use the map of symmetries from the lattice to the continuum to relate the various twisted problems on the lattice and the continuum, and then match the projective phases.

The boundary conditions on the lattice are given by $\langle g_1,g_2\rangle =\scG$, where
\ie \label{eq:lattcondgroup}
g_I \psi_{\vec{\ell}} g_I^{-1} = \psi_{\vec{\ell}}\,,\qquad I = 1,2\,.
\fe
Using the map from the lattice symmetry $G$ to the continuum symmetry $K$ given by \eqref{eq:transmap}, \eqref{eq:refomegamap} and \eqref{eq:rotmap}, we find the group elements for the continuum boundary conditions $g_I\to k_I$,
\ie \label{eq:contcondgroup}
k_I \Psi_\alpha k_I^{-1} = \Psi_\alpha\,,\qquad I = 1,2\,,
\fe
leading to $\scK = \langle k_1,k_2\rangle$.  Clearly, this map can be changed by conjugation $k \in K$, $\Psi_\alpha' = k \Psi_\alpha k^{-1}$, leading to $\scK' = \langle kk_1k^{-1},kk_2k^{-1}\rangle$.

Alternatively, as we said in the continuum (Section \ref{symmetryoftwistedcon}) and lattice (Section \ref{sec:finitelattice}) discussions, it is convenient to take a multiple cover of our continuum system with $\scK$ and the lattice system with $\scG$, such that they are untwisted and have a larger symmetry.  Then, we proceed to the problem of interest with $\scK$ and $\scG$ using a quotient by a finite group.

This group theory analysis of the twists might seem too abstract.  A more standard approach proceeds by writing explicit expressions in position space satisfying the lattice boundary conditions \eqref{eq:lattcondgroup} and the continuum boundary conditions \eqref{eq:contcondgroup} for all possible twists.  Then, a Fourier decomposition diagonalizes the lattice and continuum Hamiltonians, leading to the spectra for all possible twists.

Matching the low-energy lattice spectrum with the continuum spectrum then confirms our identification of the symmetries and the twists.  

Corresponding lattice and continuum models have the same fermion zero modes. In Table \ref{tab:twistmatch}, we list the cases with zero modes. Note that the lattice models have more inequivalent cases than the continuum models because in the continuum, there are more possible conjugations in $K$.  These zero modes lead to projective phases in the symmetry algebra. With the match of the twists and the remaining symmetry operators, we are pleased to see that the projective phases of the lattice algebra in Section  \ref{sec:latticeanomaly} match the projective phases of the continuum algebra in Section \ref{sec:paritanomaly}. This information is summarized in Table \ref{tab:latcontmatch}.
\begin{table}[t]
  \centering
  \begin{tabular}{|c|c|c|c|}
    \hline
    \multicolumn{2}{|c|}{Lattice} & \multicolumn{2}{c|}{Continuum}\\
    \hline
    Spatial Twist & Anomaly & Spatial Twist & Anomaly \\
    \hline
    $\left(1,1\right)$  & \eqref{eq:phaselat11} & $\left(1,1\right)$ & \eqref{eq:phasecont11} \\
    \hline
    $\left(1, T_2\right)$ & \eqref{eq:phaselat1g} & $\left(1,\Gamma\right)$ & \eqref{eq:phasecont1g}\\
    \hline
    $\left(1, T_2\R\right)$ & \eqref{eq:phaselat1r} & $\left(1, \sR\right)$ & \eqref{eq:phasecont1r}\\
    \hline
    $( T_1, T_2\R)$  & \eqref{eq:phaselatgrA} and \eqref{eq:phaselatgrB} & $\left(\Gamma, \sR\right)$ & \eqref{eq:phasecontgrA} and \eqref{eq:phasecontgrB}\\
    \hline
    $\left(1,  T_2 ( T_2\R)\right)$ & \eqref{eq:phaselat1gr1} & \multirow{3}{*}{$\left(1, \Gamma\sR \right)$} & \multirow{3}{*}{ \eqref{eq:phasecont1gr}}\\
    \cline{1-2}
    $\left(1,  T_1 ( T_2\R)\right)$ & \eqref{eq:phaselat1gr2} &&\\
    \cline{1-2}
    $\left(1, \C\R\right)$ & \eqref{eq:phaselat1gr3} &&\\
    \hline
\end{tabular}
  \caption{Projective phases of the continuum symmetry for various spatial twists. The rows of this table correspond to the rows in Table \ref{tab:twistmatch}.}
  \label{tab:latcontmatch}
\end{table}

\section{Conclusions} \label{Conclusions}
The parity anomaly of the continuum Dirac fermion in 2+1d can be detected by coupling the system to an  $O(2)$ background gauge field in an arbitrary 3d Euclidean non-orientable spacetime manifold.  Then, the anomaly means that the phase of the partition function is ambiguous \cite{Niemi:1983rq,Redlich:1983kn,Redlich:1983dv,Alvarez-Gaume:1984zst,Rao:1986ba, Witten:2015aba,Witten:2016cio}.

Following \cite{Delmastro:2021xox}, in this paper, we studied this anomaly in a Hilbert-space formulation by placing the theory on a spatial torus or a Klein bottle.  Starting with the theory on $\mathbb{R}^2$ with symmetry group $K$, these flat manifolds can be obtained by performing a quotient of $\mathbb{R}^2$ by an infinite discrete group $\mathscr{K}  \subset K$. The symmetry that survives in the finite-volume Hilbert space is then 
\ie
K_{\text {compact}}={N_K(\mathscr{K}) \over \mathscr{K}}\,,
\fe
i.e., transformations that preserve the entire twist data, modulo redundancies generated by the twists themselves. See Section \ref{sec:2+1dcont}, for more details.

Importantly, the quantization of the fermion zero modes on this compact space leads to a projective representation of $K_{\text {compact}}$, which probes the parity anomaly.  This analysis detects a modulo 8 anomaly.  Since this framework is limited, it cannot detect more subtle anomalies, such as the modulo 16 anomaly in \cite{Witten:2015aba,Witten:2016cio,Wang:2016qkb,Tachikawa:2016cha,Tachikawa:2016nmo}.  (See the discussion in Section \ref{sec:paritanomalyconclusion}.)

The reason we followed this Hamiltonian approach in flat space is that it can be adapted to a lattice Hamiltonian system.  The corresponding lattice system is a 2+1d staggered Majorana fermion \cite{Kogut:1974ag, Affleck:2017ubr}.  We repeated these steps on the lattice and explored the implications of various boundary conditions. Again, we started with an infinite lattice with a global symmetry group $G$, and turned it into a finite lattice using a quotient by $\scG\subset G$. As in the continuum discussion, the global symmetry of the resulting model is
\ie
G_{\text{compact}}={N_G(\mathscr{G})\over \mathscr{G}}\,.
\fe
See Section \ref{sec:staggeredfermions} for more details.

Comparing the lattice models and the continuum models, we mapped the symmetry transformations
\ie\label{symmemaps}
\begin{aligned}
T_{1,2}  &\mapsto \Gamma_{1,2}\,, \\
\mathcal{R} & \mapsto \Gamma_2\sR\,, \\
\mathcal{C}  &\mapsto e^{i \frac{\pi}{4} \sJ} e^{i \frac{\pi}{2} \sL}\,, \\
\Omega & \mapsto \Xi\,.
\end{aligned}
\fe
Here, the operators on the left-hand side are exact lattice symmetry operators, and correspondingly, the operators on the right-hand side appear as exact continuum symmetry operators.  More precisely, the spectrum of the lattice model can include states with energy of order the inverse lattice spacing $a$, which do not have well-defined quantum numbers under these continuum symmetry operators.  However, this issue does not affect the low-energy continuum theory by any relevant or irrelevant operator, and therefore, the continuum operators on the right-hand side appear as exact symmetries of the low-energy theory \cite{Cheng:2022sgb}.  The only operators here that are not such exact continuum symmetries are the individual factors in $ e^{i\frac{\pi}{4}\sJ}e^{i\frac{\pi}{2}\sL}$. See Section \ref{sec:tocontinuum} for more details.

The operator map \eqref{symmemaps} enabled us to map the spatial manifolds, the boundary conditions, the remaining symmetries, and the projective phases, thereby demonstrating the 't Hooft anomaly matching conditions.

We can summarize our results as follows.  In the continuum, we discussed an anomaly involving an interplay between the internal symmetry $\Gamma$, spatial reflection $\sR$, and fermion parity $(-1)^F$. (For simplicity, this summary ignores the full continuous internal $U(1)$ symmetry and the full continuous spatial rotation symmetry.) The anomaly was identified by considering the theory on compact spatial manifolds with various twists and finding projective representations of the symmetry algebra.  This procedure distinguishes between spatial reflection $\sR$ and time-reversal $\sT$ (or equivalently, $\Xi$).  It is also not sensitive enough to expose the whole modulo 16 anomaly of \cite{Witten:2015aba,Witten:2016cio,Wang:2016qkb,Tachikawa:2016cha,Tachikawa:2016nmo}.
\begin{itemize}
    \item Starting with a torus with periodic boundary conditions, the algebra of the symmetry operators $\Gamma$, $\sR$, and $\Xi$ \eqref{eq:phasecont11} exhibits order $2$ projective phases.
    \item Twisting by $\Gamma$, the projective representation \eqref{eq:phasecont1g} exhibits order $4$ phases, implying that for odd $N_f$, the reflection transformation $\sR$ and the anti-unitary symmetry $\Xi$ are fermionic, i.e., they anti-commute with $(-1)^F$.
    \item Twisting by $\sR$, i.e., studying the Klein bottle, the projective representation \eqref{eq:phasecont1r} exhibits order $4$ phases, implying that for odd $N_f$, the internal symmetry transformation $\Gamma$ and the anti-unitary symmetry $\Xi$ are fermionic.
     \item Twisting one cycle by $\Gamma$ and the other by $\sR$, for odd $N_f$,  $(-1)^F$ is not a standard symmetry.  And even for even $N_f$, when it is a symmetry, there are projective phases including $\Xi$ \eqref{eq:phasecontgrA} and the translation symmetry \eqref{eq:phasecontgrB}.  These phases are of order $8$.
     \item Twisting by $\Gamma\sR$, the projective algebra \eqref{eq:phasecont1gr} is of order 4.  In this case, for odd $N_f$, all of $\Gamma$, $\sR$ and $\Xi$ are fermionic.
\end{itemize}

On the lattice, the symmetry algebra is different, and the spatial reflections and time-reversal are not related.  Yet, using \eqref{symmemaps}, we find the same symmetry and the same anomaly. This is a nontrivial manifestation of 't Hooft anomaly matching.

\section*{Acknowledgments}
We are grateful to Tom Banks, Maissam Barkeshli, Simon Catterall, Meng Cheng, Yichul Choi, Davide Gaiotto, Ryohei Kobayashi, Zohar Komargodski, Sal Pace, Abhinav Prem, Brandon Rayhaun, Sahand Seifnashri, Shu-Heng Shao, Nikita Sopenko, Yuji Tachikawa, Yifan Wang, and Michael Zaletel 
for interesting discussions.  We also thank Zohar Komargodski, Shu-Heng Shao, and Yuji Tachikawa for comments about the manuscript. The work of NS was supported in part by DOE grant DE-SC0009988 and by the Simons Collaboration on Ultra-Quantum Matter, which is a grant from the Simons Foundation (Grant No. 651444). WZ was supported in part by the
Princeton University Department of Physics. 

\appendix

\section{Review of lattice fermions}
\label{app:revlat}
\subsection{From naive fermion to staggered fermion}
In this appendix, we revisit the modulated (i.e., position-dependent) internal symmetry \cite{Sala:2021jca, Han:2023fas, Pace:2024tgk} of the naive lattice fermions. By making this structure explicit and tracking its interplay with the crystalline symmetry, we will demonstrate how the naive fermions decompose into staggered fermions and derive their symmetry algebra. 

The naive fermion model on the square lattice of size $L_1 \times L_2$ with periodic boundary conditions has the Hamiltonian
\ie
  H' = i \sum_{\vec{\ell}} \sum_{\mu=1,2} \chi_{\vec{\ell}+\hat{\mu}}^{T}\,\gamma^0 \gamma^\mu\, \chi_{\vec{\ell}} \,.
\fe
where $\chi_{\vec{\ell}}$ is a real spinor. For our purposes, we assume that $L_1=L_2$ is even. Then the naive fermion enjoys an internal modulated $O(2)$ symmetry generated by $\hat C_\theta$ and $\hat K$ (the hatted operators are the symmetries of the naive fermion model),
\ie \label{eq:modO2}
  \hat C_\theta \,\chi_{\vec{\ell}}\, \hat C_\theta^{-1}
  &= \left(\cos\theta\, I + (-1)^{\ell^1+\ell^2+1}\sin\theta\, \gamma_0 \right)\, \chi_{\vec{\ell}}\,,\\
  \hat K \,\chi_{\vec{\ell}}\, \hat K^{-1}
  &= (-1)^{\ell^2}\, \gamma_2\, \chi_{\vec{\ell}}\,,
\fe
with the special element $\hat C_{\pi}=(-1)^F$. It also has crystalline symmetries generated by translations $\hat T_{1,2}$, a $\tfrac{\pi}{2}$ rotation $\hat{\C}$, and a reflection $\hat \R$,
\ie
  \hat T_i \chi_{\vec{\ell}} \hat T_i^{-1}=\chi_{\vec{\ell}+\hat{i}}\,, \qquad
  \hat{\C} \chi_{\vec{\ell}} \hat{\C}^{-1}=e^{\frac{\pi}{4}\gamma_0}\chi_{-\ell^2,\ell^1}\,, \qquad
  \hat \R \chi_{\vec{\ell}} \hat \R^{-1}=\gamma_1\chi_{-\ell^1, \ell^2}\,.
\fe
The $\mathbb{Z}_2^K$ subgroup generated by $\hat K$ is the doubling symmetry discussed in \cite{Karsten:1980wd}. The internal symmetry satisfies the group relation
\ie
  \hat K^2=1\,,\qquad \hat C_0=\hat C_{2\pi}=1\,,\qquad
  \hat K \hat C_\theta=\hat C_{-\theta}\hat K\,,\qquad
  \hat C_\theta \hat C_{\theta'}=\hat C_{\theta+\theta'}\,,
\fe
while the crystalline symmetry obeys the group relation
\ie
  & \hat T_1^{L_1}=\hat T_2^{L_2}=1\,,\qquad
  \hat{\C}^4=(-1)^F\,,\qquad
  \hat{\C}\hat T_1 \hat{\C}^{-1}=\hat T_2\,,\qquad
  \hat{\C}\hat T_2 \hat{\C}^{-1}=\hat T_1^{-1},\qquad\\
  & \hat{\R}^2=1\,,\qquad
  \hat{\R}\hat{\C}\hat{\R}^{-1}=\hat{\C}^{-1}\,.
\fe
The mixed relations between internal and crystalline symmetries are
\ie
  &\hat T_i \hat C_\theta \hat T_i^{-1}=\hat C_{-\theta}\,,\qquad
  \hat{\R} \hat C_\theta \hat{\R}^{-1}=\hat C_{-\theta}\,,\\
  &\hat T_2 \hat K \hat T_2^{-1}=(-1)^F \hat K\,,\qquad
  \hat{\C} \hat K \hat{\C}^{-1}=\hat C_{-\frac{\pi}{2}}\hat K\,,\qquad
  \hat{\R} \hat K \hat{\R}^{-1}=(-1)^F \hat K\,.
\fe

Next, we truncate the theory. At every site $\vec{\ell}$, we keep only the linear combination of $\chi_{\vec \ell}$, which is invariant under $\hat{C}_{\frac{\pi}{2}}\hat K\in O(2)$. This leaves us with  a single copy of the staggered fermion problem, with the Hamiltonian 
\ie
  H = i \sum_{\vec{\ell}} \sum_{\mu=1,2} \eta_\mu(\vec\ell)\, \psi_{\vec{\ell}+\hat{\mu}}\psi_{\vec{\ell}}, 
  \qquad \eta_1(\vec\ell)=1,\quad \eta_2(\vec\ell)=(-1)^{\ell^1}\, ,
\fe
where $\psi_{\vec{\ell}}=\frac{1}{\sqrt{2}}\left(\chi_{\vec{\ell}}^1+(-1)^{\ell_1}\chi_{\vec{\ell}}^2\right)$. The remaining symmetry is generated by
\ie
  T_1=\hat C_{-\frac{\pi}{2}}\hat T_1,\qquad
  T_2=\hat T_2\,,\qquad
  \C=\hat C_{\frac{\pi}{4}}\hat \C\,,\qquad
  \R=\hat \R\,,\qquad
  (-1)^F=\hat C_\pi\,,
\fe
and these generators satisfy the same relations as in \eqref{eq:algebrauntwisted}, in agreement with our direct derivation in Subsection \ref{sec:presentation}.\footnote{This symmetry can be derived along the lines of \eqref{eq:latcompsym}, which depends on the normalizer.}

\subsection{Going to the Continuum}
Next, we match the symmetries of the continuum limit of the naive lattice fermions with those of the staggered fermions. Recall that we limited ourselves to $L_1=L_2=2M$ with integer $M$.  Going to Fourier space, the Brillouin zone is a parallelogram centered at the origin, satisfying $\chi_{k_1+L,k_2}=\chi_{k_1,k_2+L}=\chi_{\vec{k}}$. This diagonalizes the lattice Hamiltonian and reveals four zero modes at momenta $(0,0)$, $(M,0)$, $(0,M)$, and $(M,M)$. In the low-energy limit ($|\vec{k}|\ll M$), these modes define four continuum Majorana species:
\ie
\Psi_{\vec{k}}^1=\chi_{\vec{k}}\,, \quad 
\Psi_{\vec{k}}^2=\gamma_1 \chi_{k_1+M,k_2}\,,\quad 
\Psi_{\vec{k}}^3=\gamma_2 \chi_{k_1,k_2+M}\,,\quad 
\Psi_{\vec{k}}^4=\gamma_1\gamma_2 \chi_{k_1+M,k_2+M}\,.
\fe
The effective Hamiltonian of the modes with $|\vec k|\ll M$ is diagonal in species space, $H'=\sum_{k,\alpha} \Psi^{\alpha T}_{-\vec{k}} (\gamma^0 \gamma^\mu k_\mu) \Psi^\alpha_{\vec{k}}$.

The crystalline symmetries lead to two emanant  $\mathbb{Z}_2$ symmetries, $\hat T_i \to \hat \Sigma_i e^{2\pi i P_i}$ and $\hat \R \to \hat \Sigma_2 \hat{\sR}$, which act on the continuum species as
\ie
&\hat \Sigma_1 \Psi_{\vec{k}}^{1,3} \hat \Sigma_1^{-1}= \Psi_{\vec{k}}^{1,3}\,, 
\quad 
\hat \Sigma_1 \Psi_{\vec{k}}^{2,4} \hat \Sigma_1^{-1}=- \Psi_{\vec{k}}^{2,4}\,,\\
& \hat \Sigma_2 \Psi_{\vec{k}}^{1,2} \hat \Sigma_2^{-1}=\Psi_{\vec{k}}^{1,2}\,, 
\quad 
\hat \Sigma_2 \Psi_{\vec{k}}^{3,4} \hat \Sigma_2^{-1}=-\Psi_{\vec{k}}^{3,4}\,.
\fe
and $\hat{\sR}$ is the continuum reflection along direction 1.
These generate a $\mathbb{Z}_2\times \mathbb{Z}_2$ group. The modulated $O(2)$ symmetry \eqref{eq:modO2} act as
\ie
&\hat K \Psi_{\vec{k}}^{1,2} \hat K^{-1}  = \Psi_{\vec{k}}^{3,4}\,, \qquad \hat K \Psi_{\vec{k}}^{3,4} \hat K^{-1}  = \Psi_{\vec{k}}^{1,2}\,,\\
& \hat C_\theta \Psi_{\vec{k}}^{\alpha} \hat C_\theta^{-1} 
= (A_{\theta})^\alpha_\beta \Psi_{\vec{k}}^{\beta}\,,\qquad
A_{\theta}=\cos\theta \,I - \sin \theta \,(i\sigma_2 \otimes \sigma_1)\,.
\fe
The lattice $O(2)$ symmetry and the emanant symmetry satisfy
\ie
\hat K \hat \Sigma_2 \hat K^{-1}=(-1)^F \hat \Sigma_2, \quad 
\hat \Sigma_i \hat C_\theta \hat \Sigma_i^{-1}=\hat C_{-\theta}\,.
\fe

As we said above, the lattice staggered fermions are a subset of the naive lattice fermions.  Their continuum limit is the linear combinations invariant under $\hat C_{\frac{\pi}{2}} \hat K$,
\ie
\tilde{\Psi}_{\vec{k}}^1=\Psi_{\vec{k}}^1+\Psi_{\vec{k}}^2\,, 
\quad 
\tilde{\Psi}_{\vec{k}}^2=\Psi_{\vec{k}}^3-\Psi_{\vec{k}}^4\,.
\fe
The resulting symmetry of the continuum limit of the staggered fermions that originate from the lattice (either as internal symmetries or as emanant symmetries) is $D_4$, which is generated by
\ie
\Gamma_1=\hat C_{-\frac{\pi}{2}}\hat \Sigma_1\,, \quad 
\Gamma_2= \hat \Sigma_2\,,
\fe
satisfying $\Gamma_i^2=1$ and $\Gamma_1\Gamma_2=(-1)^F \Gamma_2\Gamma_1$. This is consistent with our discussion in Subsection \ref{sec:untwistedcontinuum}.

\section{Internal symmetries of the model on a compact manifold}
\label{app:internal}

In this appendix, we derive the internal symmetries remaining after compactification from the infinite plane.\footnote{Here, we will use the notation of the continuum discussion, but the group theory is essentially the same on the lattice.} Recall that $K$ is the symmetry group of the model on the infinite plane, and we identify points using an appropriate subgroup $\scK$. Part of the full symmetry does not act on the coordinates; hence, it is an internal symmetry. It forms a normal subgroup $K_\text{int}$ of $K$. We want to find the transformations in $K_\text{int}$ that normalizes $\scK$, i.e. $N_{K_\text{int}}(\scK) \subset N_{K}(\scK)$, and then quotient out by the identifications $\scK$. 

Since $\scK$ identifies different points by the action on the coordinates, we know that $\scK \cap K_\text{int} = \{1\}$. Then we have\footnote{Recall the definition of the normalizer
\ie
N_{K_\text{int}}(\scK) = \{h \in K_\text{int}| h \scK h^{-1} = \scK \}\,,
\fe 
and the centralizer
\ie 
C_{K_\text{int}}(\scK) = \{h \in K_\text{int}| h k h^{-1} = k,\forall k \in \scK \}\,.
\fe 
}
\ie 
 N_{K_\text{int}}(\scK) = C_{K_\text{int}}(\scK)\,.
\fe  

To prove this identity, take $h \in N_{K_\text{int}}(\scK)$. For any $k \in \scK$, $h k h^{-1} \in \scK$ and hence $h k h^{-1}k^{-1} \in \scK$. On the other hand since $h \in N_{K_\text{int}}(\scK)\subset  K_\text{int}$ and $K_\text{int}$ is normalized by $K$, $k h^{-1}k^{-1} \in K_\text{int}$ and $h k h^{-1}k^{-1} \in K_\text{int}$. Therefore
\ie
hk h^{-1}k^{-1} \in \scK \cap K_\text{int} = \{1\}\,,
\fe 
which means that $h kh^{-1}=k$ and hence $h \in C_{K_\text{int}}(\scK)$. Thus, we proved $N_{K_\text{int}}(\scK) \subset C_{K_\text{int}}(\scK)$. We always have $C_{K_\text{int}}(\scK) \subset N_{K_\text{int}}(\scK)$. Therefore, the equality is proved.

To find the internal symmetry, we need to take the quotient by the identification $\scK$. Consider the quotient map $\phi_\scK: N_K(\scK) \to N_K(\scK)/\scK$. With $\scK \cap K_\text{int} = \{1\}$, the internal symmetry on the compact manifold is
\ie\label{KcurlyK} 
K_\text{int,comp} = \phi_\scK(N_{K_\text{int}}(\scK)) = N_{K_\text{int}}(\scK)=C_{K_\text{int}}(\scK)\,.
\fe 

It is standard to consider a given manifold and to examine various internal symmetry twists.  Then, the resulting unbroken internal symmetry $K_\text{int,comp}$ is given by the centralizer of the subgroup generated by the internal symmetry twist, which we can call $\scK_\text{int}$, in the full internal symmetry $K_\text{int}$, i.e., by $C_{K_\text{int}}(\scK_\text{int})$.  This is very similar to \eqref{KcurlyK}, but needs further discussion.

Let us assume that  $K = \frac{K_\text{int}\times K_\text{space}}{H}$ and $H$ is a subgroup of the center of $K$.  Then, we consider the  two following cases:
\begin{itemize}
    \item When $K$ is a direct product of the spatial symmetry and the internal symmetry, i.e., $H$ is trivial, $\scK_\text{int}\subset K_\text{int}$ is unambiguous, and therefore $C_{K_\text{int}}(\scK) = C_{K_\text{int}}\left(\scK_\text{int}\right)$.
    \item If $K$ does not factorize, i.e., $H$ is nontrivial, more care is needed.  An example is the theory of free fermions, where $H=\bZ_2^F$.  Since $\scK$ does not factorize, even though there is no ambiguity in the internal global symmetry $K_\text{int}$, the internal symmetry part of the identification,  $\scK_\text{int}$, is ambiguous.  However, since $\bZ_2^F$ is in the center of $K$, we still have $C_{K_\text{int}}(\scK)=C_{K_\text{int}}(\phi(\scK))$ where $\phi$ is the quotient map $K \to K/K_\mathrm{space}$.
\end{itemize}

\section{Twisted models and their remaining symmetries}
\label{app:decode}

\subsection{Continuum models}
\label{app:decodecont}
We now analyze \eqref{eq:contsym} in detail, reformulating it to identify the remaining symmetries. 

\bigskip\bigskip\centerline{\it Automorphism of $\scK$ and geometric transformations} \bigskip

We first study the normalizer $N_K(\scK)$ in terms of the generators of $\scK = \langle k_1,k_2\rangle$. For any $h \in N_K(\scK)$, conjugation by $h$ induces an automorphism of $\scK$. Specifically, there exists a unique matrix $X^I_J \in \mathrm{GL}(2,\bZ)$ such that
\ie \label{eq:autrep}
hk_Jh^{-1} = k_1^{X^1_J}k_2^{X^2_J}\,.
\fe 
This defines a natural homomorphism $N_K(\scK) \to \mathrm{GL}(2,\bZ)$ that gives $X(h)$. The symmetry $h$ acts on the cycles, recombining them via $X$. In particular, for an internal symmetry $h$, we have $X(h)=\delta^I_J$; while for a spatial symmetry $h$, the matrix $X$ is obtained by conjugating the representation of the symmetry operator $U(h)^T$ on the coordinates \eqref{simplT2K2}
by the shear matrix $\cs^j_J$ of \eqref{eq:Ulabel},
\ie \label{eq:autXij}
X(h)^I_J = [\cs^{-1}]_i^I[U(h)^T]^i_j\cs^j_J  \,.
\fe

As a necessary condition, the symmetry must preserve the geometry of the background manifold.  For example, the full translation group of the torus $\bR^2$ respects the shape. For rotation and reflection, we apply the automorphism argument in \eqref{eq:autXij}, which requires $X^I_J$ to be integers. For the Klein bottle, taking the reduced shape specified in \eqref{simplT2K2}, the symmetric translation group is $\bZ_2 \times U(1)$, generated by $e^{i \pi \cl_1P_1}$ and $e^{2\pi i a^2P_2}$. And while continuous rotations $e^{i \theta \sL}$ are generally not a symmetry, reflections $\sR$ and $\sR e^{i \pi \sL}$ may still preserve the geometry.

\bigskip\bigskip\centerline{\it Internal and reflection twists} \bigskip

We are primarily interested in the symmetries that do not involve continuous translations.  Therefore, we consider the quotient $\varphi_\text{space}: K \to K/\bR^2 = \bar{K}$. We have
\ie \label{eq:wotransbound}
\varphi_\text{space}(N_K(\scK)) \subset N_{\bar{K}}\left(\varphi_\text{space}(\scK))\right)\,.
\fe 
This implies that the remaining symmetry must respect the internal symmetry twists and reflection twists.
For models with fermion zero modes (discussed in Appendix \ref{app:twistedcontinuum}), one can explicitly verify, using the set of remaining symmetries listed in Subsection \ref{sec:paritanomaly}, that there always exists a fundamental domain (a ``good shape'') such that the symmetries extend to the maximal set $\varphi_\text{space}(N_K(\scK)) = N_{\bar{K}}\left(\varphi_\text{space}(\scK)\right)$. Thus, we can label the models by $\varphi_\text{space}(\scK)$ for analyzing the anomalies.

For cleaner notation, we introduce the translations along the two cycles of the fundamental domain (a special case was introduced in Subsection \ref{sec:paritanomaly})
\ie \label{eq:cycleP}
\bar P_I = \frac{1}{2\pi}\cs^i_I P_i\,.
\fe
The boundary conditions are then written as
\ie \label{eq:contbc}
e^{2\pi i \bar P_I} = \bar k_I\,,
\fe 
with $\bar k_I \in K/\bR^2$. 

\subsection{Lattice models}
\subsubsection{Lattice twists from fundamental groups}
\label{app:fundgroup}

In this appendix, we identify an overcomplete set of pairs of $g_{1,2}$ that generate the fundamental group of the manifold and specify the lattice twists. This requires that $g_{1,2}$ have infinite order and satisfy the relations
\ie \label{eq:fundrelation}
g_1g_2= g_2g_1\,,\qquad \text{ or }\qquad g_1g_2= g_2^{-1}g_1\,
\fe 
for the torus or the Klein bottle, respectively.
In general, any $g_I \in G$ can be written as 
\ie\label{generatorg} 
g_I = T_1^{S^1_I}T_2^{S^2_I}\R^{m_I}\C^{n_I}[(-1)^F]^{W_I}\,,
\fe 
where $S^i_I\in \bZ$, $n_I=0,1,2,3$, and  $m_I, W_I=0,1$ (if $n_I \neq 0$, then $m_I\ne 0$; otherwise, $g_I$ has a finite order). We will denote
\ie 
T_{\vec{S}_I} = T_1^{S^1_I}  T_2^{S^2_I}\,, \qquad   \R_{\theta_I} = \C^{n_I}\R  \,,
\fe 
where $\R_{\theta_I}$ is the reflection of the line with angle $\theta_I=\frac{n_I\pi}{4}$. Then, the generator \eqref{generatorg} is
\ie 
g_I = T_{\vec{S}_I}\R_{\theta_I}^{m_I}[(-1)^F]^{W_I}\,.
\fe 

We first focus on the geometry and quotient by the internal symmetry $(-1)^F$, i.e., $G \to G/\bZ_2^F$. The projections $\tilde g_{1,2} \in G/\bZ_2^F$ should generate the corresponding fundamental groups. The requirement that $\bZ^2$ is the universal cover of the torus ($\Ts$) or the Klein bottle ($\Kb$) constrains the shape parameters $\vec{S}_I$ and $m_I$ as follows.
\begin{enumerate}[(a)]
    \item For the torus, we require $m_1=m_2=0$. We can recombine the cycles such that $S^2_1=0$. Written explicitly, the generators can be chosen as
    \ie \label{eq:shapeTs}
    \tilde g_1 = \tilde T_1^{L_1}\,,\quad \tilde g_2 = \tilde T_1^b \tilde T_2^{L_2}
    \fe 
    with integers $L_I$ and $b$, such that $0 \leq b < L_1$ and $0<L_2$.\footnote{Note that in this case $L_2 \neq 0$ otherwise $\langle \tilde g_1,\tilde g_2\rangle \cong \langle \tilde T_1^{\mathrm{gcd}(L_1,b)}\rangle$ is a finite group.}
    \item For the Klein bottle, we require one of the $m_I$ to be 0. Setting $m_1=0$, we impose the relation
    \ie 
    \tilde{\R}_{\theta_2} \tilde T_{\vec{S}_1} \tilde{\R}_{\theta_2}^{-1} = \tilde T_{-\vec{S}_1}\,.
    \fe 
    This yields the relation $\tan \theta_I = S_1^2/S_1^1$. Given $\theta_I=\frac{n_I\pi}{4}$, we find
    \ie 
    S_1^2=\pm S_1^1\,, \quad \text{ or } \quad S_1^1=0\,, \quad \text{ or } \quad S_1^2=0\,.
    \fe 
    We can choose $\theta_2=0$ or $\theta_2=\pi/4$ (others are equivalent by conjugation by $\tilde \C$), resulting in two cases, whose generators can be chosen as
    \ie \label{eq:shapeKbj}
    &\tilde g_1 = \tilde T_1^{L_1}\,, \quad \tilde g_2 = \tilde T_1^{S^1_2} \tilde T_2^{S^2_2}\tilde \R \qquad , \qquad S^2_2>0\,,\\
    &\tilde g_1 = \tilde T_1^{L_1} \tilde T_2^{L_1}\,, \quad  \tilde g_2 = \tilde T_1^{S^1_2} \tilde T_2^{S^2_2}\tilde \C \tilde \R\qquad, \qquad S^1_2>S^2_2\,,
    \fe
    Finally, conjugating by $\tilde T_1^n$ allows us to further reduce the inequivalent cases to
    \ie \label{eq:shapeKb}
    & \tilde g_1=\tilde  T_1^{L_1}\,, \quad \tilde g_2=\tilde T_2^{L_2} \tilde \R\,,\\
    & \tilde  g_1=\tilde T_1^{L_1}\,, \quad \tilde g_2=\tilde T_2^{L_2} (\tilde T_1\tilde \R)\,,\\
    & \tilde g_1=\tilde T_1^{L_1}\tilde T_2^{L_1}\,, \quad \tilde g_2=\tilde T_1^{L_2} \tilde T_2^{-L_2}(\tilde \C \tilde \R)\,,\\
    & \tilde g_1=\tilde T_1^{L_1}\tilde T_2^{L_1}\,, \quad \tilde g_2=\tilde T_1^{L_2}\tilde T_2^{-L_2}(\tilde T_2\tilde \C \tilde \R)\,.
    \fe
     with integers $L_I$ such that $0<L_I$ for the first three cases and  $0<L_1\,, 0\leq L_2$ for the last case.\footnote{In the first three cases, $L_I \neq 0$ otherwise the order of $\tilde g_I$ is finite.}
\end{enumerate} 
We then consider the full group $G$ including $(-1)^F$. The relation \eqref{eq:fundrelation} constrains $L_{1,2}$ in \eqref{eq:shapeTs} and \eqref{eq:shapeKb} (see Footnotes \ref{ft:shape1}--\ref{ft:shape3} for details). We may also introduce $(-1)^F$ twists into \eqref{eq:shapeTs} and \eqref{eq:shapeKb}$;$ some choices are equivalent up to conjugation.

\subsubsection{Lattice symmetry and structure}

Here, we discuss the symmetry of the compact lattice model \eqref{eq:latcompsym}, structuring the discussion in parallel to the continuum case in \ref{app:decodecont}. 

First, as in the continuum, the symmetry transformations should respect the shape of the finite lattice.

Next, to mimic the quotient $K/\bR^2$ in the continuum, we consider the normal subgroup $\scT=\langle T_1^2, T_2^2\rangle$ of $G$.  The reason we focus on this subgroup is that, as we will see, models that differ by twists by $\scT$ have similar symmetries and similar fermion zero modes. Then, we take the quotient $\varphi_\mathrm{crys}: G \to \bar{G} = G/\scT$.\footnote{The quotient group
\ie
\bar{G} = G/\scT \cong \left((\bZ_4 \times \bZ_4) \rtimes (\bZ_2 \times \bZ_2)\right) \times \bZ_2\,,
\fe
is generated by $\bar{T}_1\bar{T}_2$, $\bar{\C}$, $ \bar{\R}\bar{T}_2$, $\bar T_1$, $ \bar \Omega$.
The nontrivial group relations for these generators (suppressing elements that commute) are
\ie
& \left(\bar{T}_1\bar{T}_2\right)^4=\bar{\C}^4=\left(\bar{\R}\bar{T}_2\right)^2=\bar\Omega^2=\bar T_1^2=1\,,\qquad  \bar T_1 \left(\bar{T}_1\bar{T}_2\right) \bar T_1^{-1} =  \left(\bar{T}_1\bar{T}_2\right)^{-1}\,,\\
& \bar T_1 \bar \C \bar T_1^{-1} =  \left(\bar{T}_1\bar{T}_2\right)\bar{\C}\,,\qquad \left(\bar{\R}\bar{T}_2\right) \bar \C \left(\bar{\R}\bar{T}_2\right)^{-1} =  \left(\bar{T}_1\bar{T}_2\right) \bar{\C}^{-1}\,,
\fe
where $(-1)^F = (\bar T_1 \bar T_2)^2$. This group also gives the exact symmetries in the continuum of the lattice staggered model, as discussed in Subsection \ref{sec:untwistedcontinuum}.} 
Similar to \eqref{eq:wotransbound}, we have
\ie \label{eq:lattice-wotransbound}
\varphi_\mathrm{crys}\left(N_G(\scG)\right) \subset N_{\bar G}\!\left(\varphi_\mathrm{crys}(\scG)\right)\,.
\fe
In more detail, changing the twist by an element of $\scT$ can change the symmetries of the model (e.g., it might not be rotation invariant).  But as in the discussion following \eqref{eq:wotransbound},  for models with zero modes, there exists a convenient choice of twist and fundamental domain (a ``good shape'') which is maximally symmetric, i.e.,
\ie
\varphi_\mathrm{crys}\left(N_G(\scG)\right)=N_{\bar G}\!\left(\varphi_\mathrm{crys}(\scG)\right)\,.
\fe
Then, we will label this maximally symmetric model by $\varphi_\mathrm{crys}(\scG)$. (These maximally symmetric models will give us the best diagnostic of the anomalies.)  This means that we only write the twist with the power of $T_{1,2}$ mod 2.

\section{Review of continuum fermions on a Klein bottle}
\label{app:kleinfermioncont}

In this appendix, we review the boundary conditions, symmetries, and spectrum of free continuum fermions on a spatial Klein bottle, as introduced in Section \ref{sec:2+1dcont}. See related discussion in \cite{Hsieh:2015xaa}. This appendix prepares us for the solution of the lattice models on the Klein bottle and their matching with the continuum in Appendix \ref{app:twistedcontinuum}.

\subsection{Majorana fermion}
\label{app:kelinreal}
Consider a free continuum Majorana fermion with Hamiltonian density $\mathcal{H}=i \Psi^{T} \gamma^0 \gamma^j \partial_j \Psi$, defined on the fundamental domain $\cl_1 \times \cl_2$ with twisted boundary conditions\footnote{Here, we do not include twists by $(-1)^F$. Such twists can be added easily.}
\ie
  \psi(x^1+\cl_1,x^2)=\psi(x^1,x^2)\,, \quad \psi(-x^1,x^2+\cl_2)=\gamma_1\psi(x^1,x^2)\,.
\fe
These conditions correspond to a quotient of the universal cover $\bR^2$ by the group $\scK$ generated by the translation $e^{ i \cl_1P_1}$ and the glide $e^{ i \cl_2P_2}{\sR}$. As in Subsection \ref{sec:paritanomaly}, we rescale $\bar{P}_1=\frac{\cl_1}{2\pi}P_1$ and $\bar{P}_2=\frac{\cl_2}{2\pi}P_2$ and impose 
\ie\label{KBconditA}
e^{ 2\pi i \bar P_1}= e^{2\pi i \bar P_2}{\sR}=1\,.
\fe

To determine the spectrum, we consider the double cover of the Klein bottle, namely a torus $\cl_1 \times 2\cl_2$ with periodic boundary conditions.
We expand in Fourier modes on this torus to diagonalize the Hamiltonian and then use the Fourier-transformed twisted boundary condition to identify the modes that reduce to the Klein bottle.

The resulting Fourier decomposition on the Klein bottle is
\ie 
\psi(x^1,x^2) & = \sum_{k_1 \in \bZ^+\,, \ k_2 \in \tfrac{\bZ}{2}} e^{i\frac{2\pi}{\cl_2}k_2x^2}\left(e^{i\frac{2\pi}{\cl^1}k_1x^1}+e^{-i\frac{2\pi}{\cl^1}k_1x^1}(-1)^{2k_2}\gamma_1\right)\psi_{k_1,k_2}\\
&\qquad + \sum_{k_2\in \tfrac{\bZ}{2}} e^{i\frac{2\pi}{\cl_2}k_2x^2}\psi_{0,k_2}\,,\\
\psi_{0,k_2 }&=(-1)^{2k_2}\gamma_1\psi_{0,k_2}\,, \\
\psi_{k_1,k_2}^{\dagger} &= (-1)^{2k_2}\gamma_1 \psi_{k_1,-k_2}\quad (k_1>0)\,,\qquad 
\psi_{0,k_2}^{\dagger} = \psi_{0,-k_2}\quad (k_1= 0)\,.
\fe 
Here, $k_2$ reflects the translation symmetry on the Klein bottle, while $k_1$  reflects the symmetry from the double cover.

For $k_1>0$, $\psi_{\vec{k}}$ is a two-component spinor, and for $k_1=0$, the two-component $\psi_{\vec{k}}$ is constrained, leading to a single real degree of freedom. 
The diagonalized Hamiltonian is 
\ie
  &H=\sum_{k_1 \in \bZ^+\,,\ k_2 \in \tfrac{\bZ}{2}} \psi_{\vec{k}}^\dagger D(\vec{k})\psi_{\vec{k}}+\sum_{k_2 \in \tfrac{\bZ}{2}}(-1)^{2k_2}\frac{2\pi}{\cl_2}k_2\psi_{0,k_2}^\dagger\psi_{0,k_2} \,,\\
&D(\vec{k})=2\pi\gamma_0\left(\gamma^1{k_1\over \cl_1}+\gamma^2{k_2\over \cl_2}\right)\,.
\fe
This leads to a standard Fock space spectrum with the single real zero mode at $(0,0)$ leading to a unique ground state.\footnote{If the reader is concerned about having an odd number of fermion zero modes, they can take two copies of the system.}

Finally, the action of the spatial symmetries on the momentum modes is
\ie \label{eq:contkleinrealaction}
   e^{i\epsilon \bar P_2}\psi_{\vec{k}}e^{-i\epsilon \bar P_2}=e^{i\epsilon k_2}\psi_{\vec{k}}\,, \quad e^{i\pi \bar P_1}\psi_{\vec{k}}e^{-i\pi \bar P_1}=(-1)^{k_1}\psi_{\vec{k}}\,, 
  \quad {\sR}\psi_{\vec{k}}{\sR}^{-1}=(-1)^{2k_2}\psi_{\vec{k}}\,. 
\fe
As a check, this reproduces \eqref{KBconditA}.

As we said above, we can insert $(-1)^F$ defects.  Doing it along cycle 1 leads to an inequivalent model.  However, doing it along cycle 2 yields a unitarily equivalent model, related by conjugation by reflection $e^{i\pi \sL}\sR$ along cycle 2.

Finally, we diagonalize the Hamiltonian $H = \sum_k E_k \!:\!a_k^\dagger a_k\!:\! + E_{\mathrm{vac}}$, where $a_k^\dagger$ are creation operators of non-negative energy states. Then the total momentum is
\ie \label{eq:shiftinf}
P_2 &= \sum_{k_1 > 0\,, k_2 \in \tfrac{\bZ}{2}} k_2 \!:\!a_k^\dagger a_k\!:+ \sum_{\substack{k_1 = 0\,,  k_2 \in \bZ^+}} k_2 \!:\!a_k^\dagger a_k\!:\! + \sum_{k_1 = 0\,,  k_2 \in \bZ^-+\frac{1}{2}} k_2 \!:\!a_k^\dagger a_k\!: + P_{\text{vac}}\\
P_{\text{vac}} &= \frac{1}{2} \left( -\sum_{\substack{k_2 \in \mathbb{Z}+\frac{1}{2}\,,\\ k_2 \leq 0}} |k_2| + \sum_{\substack{k_2 \in \mathbb{Z}\,,\\ k_2 > 0}} k_2 \right) = \frac{1}{2} \left( -\zeta\left(-1, \frac{1}{2}\right) + \zeta(-1, 0) \right) = -\frac{1}{16}.
\fe
Thus, the momentum is shifted by $-\tfrac{1}{16}$, contributing a mod 8 phase $e^{2\pi i \bar P_2}\sR=e^{-i \frac{\pi}{8}}$.

\subsection{Dirac fermion and twist of $U(1)$}
Here, we consider a free Dirac fermion $\psi$ (i.e., two Majorana fermions) on the Klein bottle with internal $U(1)$ twists
\ie
\psi(x^1+\cl_1,x^2)=e^{i\theta_1}\psi(x^1,x^2), \quad \psi(-x^1,x^2+\cl_2)=e^{i\theta_2}\gamma_1\psi(x^1,x^2)\,.
\fe
The momenta of the Fourier modes on the Klein bottle are shifted $k_1 \in \bZ + \frac{\theta_1}{2\pi}$ and $k_2 \in \frac{\bZ}{2}+\frac{\theta_2}{2\pi}$. For $\theta_1= 0$ and $\theta_2 = 0,\pi$, they correspond to the cases discussed in Appendix \ref{app:kelinreal}. For other values of $\theta_{1,2}$, there is no zero mode.

The symmetries act as
\ie
  e^{i\epsilon \bar P_2}\psi_{\vec{k}}e^{-i\epsilon \bar P_2}=e^{i\epsilon k_2}\psi_{\vec{k}}\,, \quad e^{i\pi \bar P_1}\psi_{\vec{k}}e^{-i\pi \bar P_1}=(-1)^{k_1}\psi_{\vec{k}}\,, \quad {\sR}\psi_{\vec{k}}{\sR}=e^{i\theta_2}e^{-2\pi i k_2}\psi_{\vec{k}}\,,
\fe
and they satisfy the boundary conditions
\ie
  e^{2\pi i \bar P_1}=e^{i \theta_1 \sJ}\,,\quad e^{2\pi i \bar P_2}{\sR_{1}}=e^{i\theta_2 \sJ}\,.
\fe

\section{Staggered fermion with various twists}
\label{app:twistedcontinuum}

\subsection{Untwisted staggered fermion}
\label{app:untwistedexp}
In this subsection, we study the untwisted case, i.e., the system on the torus with periodic boundary conditions including shear, as in Subsection \ref{sec:untwistedcontinuum}. We solve the spectrum, analyze the action of the lattice symmetries on the momentum modes, and finally, the projective representation of the lattice and continuum symmetries through the zero modes. The discussion of the twisted cases in the later subsection follows a similar analysis.

\subsubsection{Diagonalizing the Hamiltonian}

The boundary conditions with shear \eqref{eq:latticeshear} in position space, lead to the Fourier expansion
\ie \label{eq:fourier}
  \psi_{\vec{\ell}}=\frac{1}{\sqrt{2\det(S)}}\sum_{\vec{k} \in \mathrm{BZ}} e^{2\pi i \vec{k}^T S^{-1}\vec{\ell}} \psi_{\vec{k}}\,,
\fe
where $\psi^\dagger_{\vec{k}} =\psi_{-\vec{k}} $ and
$\psi_{\vec{k}+\vec{S}^i}=\psi_{\vec{k}}$, thus defining a parallelogram Brillouin zone ($\mathrm{BZ}$) centered at the origin.\footnote{Orthogonality of the Fourier modes follows from
\ie
\frac{1}{2\det(S)}\sum_{\vec{k} \in \mathrm{BZ}} e^{2\pi i \vec{k}^TS^{-1}(\vec{\ell}-\vec{\ell}')}=\delta_{\ell,\ell'}\,,\quad 
\frac{1}{2\det(S)}\sum_{\vec{\ell}} e^{2\pi i(\vec{k}-\vec{k}')^TS^{-1}\vec{\ell}}=\delta_{k,k'}\,.
\fe }   The Fourier modes satisfy $\{\psi_{\vec{k}},\psi_{\vec{k}'}\}=2\delta_{\vec{k},-\vec{k}'}$.

Substituting this expansion into the Hamiltonian reveals four real zero modes at $\vec{k} \in \{ \vec{0}, \frac{\overrightarrow{S^1}}{2}, \frac{\overrightarrow{S^2}}{2}, \frac{\overrightarrow{S^1}+\overrightarrow{S^2}}{2} \}$.  Expanding around these momenta, we combine the modes into a spinor $\tilde{\psi}_{\vec{k}} = (\psi_{\vec{k}}, \psi_{\vec{k}+\frac{\overrightarrow{S^1}}{2}})^T$, leading to the Hamiltonian 
\ie\label{eq:Dk}
& H = \sum_{\vec{k} \in \mathrm{BZ}'} \tilde{\psi}^\dagger_{\vec{k}} D(\vec{k})\tilde{\psi}_{\vec{k}}\,,\\
& D(\vec{k}) =  \sin(2\pi (S^{-T}\vec{k})_1)\gamma_0\gamma_1 + \sin(2\pi (S^{-T}\vec{k})_2)\gamma_0\gamma_2\,,
\fe 
and the reduced Brillouin zone $\mathrm{BZ}'$ is restricted with further identification $\tilde{\psi}_{\vec{k}} = \gamma_1 \tilde{\psi}_{\vec{k}+\frac{\overrightarrow{S^1}}{2}}$. Diagonalization yields $H= \sum_{\vec{k}} E_{\vec{k}} \, a_{\vec{k}}^\dagger a_{\vec{k}}$, with energy $E_{\vec{k}} = \left[ \sum_\mu \sin^2(2\pi (S^{-T}\vec{k})_\mu) \right]^{1/2}$. The eigenmodes are defined by
\ie
   a_{\vec{k}}=\cos \frac{\tilde{\theta}_k}{2} \tilde{\psi}_{\vec{k}}^1 + \sin \frac{\tilde{\theta}_k}{2} \tilde{\psi}_{\vec{k}}^2 \,,
\fe
where $\tilde{\theta}_k=\operatorname{Arg}\left[\sin(2\pi (S^{-T}\vec{k})_1)+i \sin(2\pi (S^{-T}\vec{k})_2)\right]$. For $\vec{k}=\vec{0}$ or $\frac{\overrightarrow{S^2}}{2}$, we take $a_{\vec{k}} = (\psi_{\vec{k}}^1+i \psi_{\vec{k}}^2)/\sqrt{2}$.

In the low-energy limit near the zero modes, the system reduces to two relativistic Majorana fermions (equivalently, one Dirac fermion)
\ie\label{eq:lowmode}
    \Psi_{\vec{k}}^1=\tilde{\psi}_{\vec{k}}, \quad \Psi_{\vec{k}}^2=\gamma_2 \tilde{\psi}_{\vec{k}+\tfrac{\overrightarrow{S^2}}{2}}\,,\quad (|\vec{k}| \ll \sqrt{\det(S)})\,.
\fe
with the Hamiltonian
\ie
  H = 2\pi \sum_{\alpha=1,2} \sum_{\vec{k}} \Psi_{-\vec{k}}^{\alpha T} \left[ \gamma^0\gamma^\mu (S^{-T}\vec{k})_\mu \right] \Psi_{\vec{k}}^\alpha\,,
\fe
recovering $E(k)=2\pi |S^{-T}\vec{k}|$.
\subsubsection{Action of the lattice symmetry on the low-energy modes}
The symmetry operators act on the momentum modes as, 
\ie 
  T_1 \psi_{\vec{k}}T_1^{-1}&=e^{2\pi i k_I(S^{-1})^I_1} \psi_{\vec{k}+\frac{\overrightarrow{S^2}}{2}}\,, \\
  T_2 \psi_{\vec{k}}T_2^{-1}&=e^{2\pi i k_I(S^{-1})^I_2}  \psi_{\vec{k}}\,, \\
    \Omega\psi_{\vec{k}}\Omega^{-1}&=\psi_{-\vec{k}}\,. 
\fe
For certain shear matrices, $\R$ or $\C$ are also symmetries, and then, we also have
\ie
{\R}\psi_{\vec{k}}{\R}^{-1}&=\psi_{X(\R)^{-T}\vec{k}+\frac{\overrightarrow{S^1}}{2}}\,,\\
\C\psi_{\vec{k}}\C^{-1}&=\frac{1}{2}\left(
        \psi_{X(\C)^{-T}\vec{k}} + \psi_{X(\C)^{-T}\vec{k}+\frac{\overrightarrow{S^1}}{2}} - \psi_{X(\C)^{-T}\vec{k}+\frac{\overrightarrow{S^2}}{2}} + \psi_{X(\C)^{-T}\vec{k}+\frac{\overrightarrow{S^1}+\overrightarrow{S^2}}{2}}
    \right)\,,\\
\fe 
where $X(\R),X(\C)\in\mathrm{GL}(2,\bZ)$, are given in \eqref{eq:autXij}. These expressions motivate the map between the lattice and the continuum symmetries in Subsection \ref{sec:untwistedcontinuum}, as summarized in \eqref{latticetocom}.\footnote{The continuum symmetries, such as $\Gamma_{1,2}$, are not exact lattice symmetries. However, since our lattice theory is free, we can extend the action of $\Gamma_{1,2}$ from the low-energy modes to the full theory, leading to exact lattice symmetries.
A similar phenomenon was discussed in the $1+1$d Majorana chain \cite{Seiberg:2025zqx}.  Such an extension of the symmetry cannot be done once the lattice action is more generic.}

\subsubsection{Projective representation of lattice and continuum symmetries}

Here, we find the projective representation of the symmetry.  The projective phases arise only from the fermion zero modes.  Therefore, it is enough to focus on them.  These zero modes are the same on the lattice and in the continuum, thus guaranteeing that the phases match.  

Let us see it in more detail.  There are four real zero modes at $\vec{k} \in \{ \vec{0}, \frac{\overrightarrow{S^1}}{2}, \frac{\overrightarrow{S^2}}{2}, \frac{\overrightarrow{S^1}+\overrightarrow{S^2}}{2}\}$. We label them as $\psi^{i=1,2,3,4}$. They are components of $\Psi$ in \eqref{eq:lowmode}.  They are represented on the ground states as
\ie 
\psi^1 = \sigma^1 \otimes I_2\,,\quad \psi^2 = \sigma^2 \otimes I_2\,,\quad \psi^3 = \sigma^3 \otimes \sigma^1 \,,\quad \psi^4= \sigma^3 \otimes \sigma^2\,.
\fe 
The continuum unitary symmetry operators are represented as
\ie 
& \Gamma = i\psi^3\psi^4\,, \quad &&\sR = \frac{i}{2}(\psi^1-\psi^2)(\psi^3-\psi^4)\,,\\ 
& \sJ = -\frac{i}{2}(\psi^1\psi^3+\psi^2\psi^4)\,, \quad && \sL = -\frac{i}{4}(\psi^1\psi^2+\psi^3\psi^4)\,.
\fe 
This leads to the projective phases
\ie 
e^{i\pi \sJ}& = - e^{2\pi i \sL}\,,\qquad \Gamma\sR &= - \sR \Gamma\,.
\fe 
Since these symmetries are unitary, taking an even number of copies cancels these two phases. Following the calculation of Appendix C of \cite{Seiberg:2025zqx}, we can also derive the projective phases involving $\Xi$ with $N_f$ copies of Dirac fermions,\footnote{Here, one needs to identify $\Gamma$ and $\sR$ as $\sC$, and identify $\Xi$ as $\Theta$ of \cite{Seiberg:2025zqx}. There is no projective phase between $\Xi$ and $e^{i\theta \sL}$ or $e^{i\theta \sJ}$ (with the normalization $e^{4\pi i \sL}=e^{2\pi i \sJ}=1$), which can be checked by conjugating them by $\Xi$.}
\ie 
\Xi \Gamma & = (-1)^{N_f} \Gamma \Xi  \,,\\
\Xi \sR & = (-1)^{N_f} \sR \Xi\,,  \\
\Xi^2 & = (-1)^{N_f}\,.
\fe 
The projective phases of the lattice symmetry operators can be derived using the same calculation by identifying $T_1 \to \Gamma e^{i\frac{\pi}{2}\sJ}$, $T_2\R \to \sR$, $\C \to e^{i\frac{\pi}{4}\sJ}e^{i\frac{\pi}{2}\sL}$ and $\Omega \to \Xi$. 
All these phases are of order 2.

\subsection{Twisted staggered fermions on the torus}
\label{app:twisttorus}

Here, we analyze the twisted staggered fermion on a torus. We follow the general twists
\ie
g_1 = T_1^{S_1^1}T_2^{S_1^2}\left[(-1)^F\right]^{W_1}\,,\quad
g_2 = T_1^{S_2^1}T_2^{S_2^2}\left[(-1)^F\right]^{W_2}\,,
\fe
and identify the cases with the same $S^i_I$ modulo 2.  We also identify cases using cycle redefinition and conjugation. These twisted cases have more general shapes than those presented in the Subsection \ref{sec:latticeanomaly}.

Table \ref{tab:torustwist} lists the inequivalent lattice twists on the torus, their zero modes, and their matching to the continuum twists.  The diagonalization of the energy spectrum follows Appendix \ref{app:untwistedexp} and the doubling trick in Appendix \ref{app:kelinreal}. The projective phases are derived as in Appendix \ref{app:untwistedexp} and in \cite{Seiberg:2023cdc,Seiberg:2025zqx}.

\begin{table}[t]
    \centering
    \begin{tabular}{|c|c|c|c|c|c|c|c|}
       \hline 
       \makecell{Lattice \\ Twist}  & \makecell{Boundary \\ Condition} & $S_1^1$ &  $S_1^2$ & $S_2^1$ &  $S_2^2$ & \makecell{Real Zero\\Modes} & \makecell{Continuum \\ Twist}\\
       \hline
       \hline
       $(1,1)$ & \makecell{$\psi_{\vec{\ell}}=\psi_{\vec{\ell}+\vec{S}_1}$\\$\psi_{\vec{\ell}}=\psi_{\vec{\ell}+\vec{S}_2}$ } & 0 & 0 & 0 & 0 & 4 & $(1,1)$\\
       \hline
       $(1,T_2)$ & \makecell{$\psi_{\vec{\ell}}=\psi_{\vec{\ell}+\vec{S}_1}$\\$\psi_{\vec{\ell}}=\psi_{\vec{\ell}+\vec{S}_2}$ } & 0 & 0 & 0 & 1 & 2 & $(1,\Gamma)$\\
       \hline
       $((-1)^F,1)$ & \makecell{$\psi_{\vec{\ell}}=-\psi_{\vec{\ell}+\vec{S}_1}$\\$\psi_{\vec{\ell}}=\psi_{\vec{\ell}+\vec{S}_2}$ } & 0 & 0 & 0 & 0 & 0 & $((-1)^F,1)$\\
       \hline
       $((-1)^F,T_2)$ & \makecell{$\psi_{\vec{\ell}}=-\psi_{\vec{\ell}+\vec{S}_1}$\\$\psi_{\vec{\ell}}=\psi_{\vec{\ell}+\vec{S}_2}$ } & 0 & 0 & 0 & 1 & 0 & $((-1)^F,\Gamma)$\\
       \hline
       $(1,T_1T_2)$ & \makecell{$\psi_{\vec{\ell}}=\psi_{\vec{\ell}+\vec{S}_1}$\\$\psi_{\vec{\ell}}=(-1)^{\ell_2}\psi_{\vec{\ell}+\vec{S}_2}$ } & 0 & 0 & 1 & 1 & 0 & $(1,e^{i\frac{\pi}{2}\sJ})$\\
       \hline
    \end{tabular}
    \caption{Five inequivalent lattice twists on the torus and their continuum counterparts. For each lattice twist, we list the induced boundary conditions along the two cycles generated by $\vec S_{1,2}$, labeled by $S_I^i \bmod 2$ (even/odd), and the number of real fermion zero modes. The last column presents the corresponding continuum twists $(\bar{k}_1,\bar k_2)$ such that $e^{2\pi i\bar P_I}=\bar k_I$ defined in \eqref{eq:contbc}.}
    \label{tab:torustwist}
\end{table}

\subsection{Twisted staggered fermions on the Klein bottle}

Here, we repeat the torus discussion in Appendix \ref{app:twisttorus} on the Klein bottle. We follow the general Klein bottle twists in \eqref{eq:shapeKbj}, namely
\ie
g_1 = T_1^{L_1}\left[(-1)^F\right]^{W_1}\,, \qquad
g_2 = T_1^{S_2^1}T_2^{S_1^2}\R\left[(-1)^F\right]^{W_2}\,,
\fe
or
\ie
g_1 = T_1^{L_1}T_2^{L_1}\left[(-1)^F\right]^{W_1}\,, \qquad
g_2 = T_1^{S_2^1}T_2^{S_1^2}\C\R\left[(-1)^F\right]^{W_2}\,,
\fe
with parameters $L_2$ and $S_2^i$ subject to the constraints $L_1(S_2^2-1)\in 2\bZ$ or $L_1(S_2^1+S_2^2)\in 2\bZ$, respectively (see also Subsection \ref{sec:latTsandKb}). We identify cases with the same $L_2, S_2^i$ modulo 2, and cases related by cycle redefinition and conjugation.

Table \ref{tab:KBtwist} lists the inequivalent lattice twists on the Klein bottle, their zero modes, and their matching with the continuum twists. The detailed derivation of the spectrum and the projective phases follows the same methods and, hence, it is again omitted.

\begin{table}[!htbp]
    \centering
    \begin{tabular}{|c|c|c|c|c|c|c|}
      \hline
      \makecell{Lattice\\Twist} & \makecell{Boundary Condition} & $L_1$ & $S_1^1$ & $S_1^2$ &  \makecell{Real Zero\\Modes} & \makecell{Continuum \\ Twist}\\
      \hline
      \hline 
      $(1,T_2\R)$ & \makecell{$\psi_{\vec{\ell}}=\psi_{\ell^1+L_1,\ell^2}$,\\$\psi_{\vec{\ell}}=(-1)^{\ell_1}\psi_{-\ell^1+S_1^1,\ell^2+S_1^1}$} & 0 & 0 & 1 & 2 & $(1,\sR)$\\
      \hline 
      $(T_1,T_2\R)$ & \makecell{$\psi_{\vec{\ell}}=(-1)^{\ell_2}\psi_{\ell^1+L_1,\ell^2}$,\\$\psi_{\vec{\ell}}=(-1)^{\ell_1}\psi_{-\ell^1+S_1^1,\ell^2+S_1^1}$} & 1 & 0 & 1 & 1 & $(\Gamma,\sR)$\\
      \hline 
      $(1,\R)$ & \makecell{$\psi_{\vec{\ell}}=\psi_{\ell^1+L_1,\ell^2}$,\\$\psi_{\vec{\ell}}=(-1)^{\ell_1}\psi_{-\ell^1+S_1^1,\ell^2+S_1^1}$} & 0 & 0 & 0 & 2 & $(1,\sR\Gamma)$\\
      \hline 
      $(1,T_1T_2\R)$ & \makecell{$\psi_{\vec{\ell}}=\psi_{\ell^1+L_1,\ell^2}$,\\$\psi_{\vec{\ell}}=f(\vec{\ell})\psi_{-\ell^1+S_1^1,\ell^2+S_1^1}$\\
      $f(\vec{\ell})=(-1)^{\ell_1+\ell_2}$}& 0& 1&1 & 2 & $(1, \sR\Gamma)$\\
    \hline 
    $(1,\C\R)$ & \makecell{$\psi_{\vec{\ell}} = \psi_{\ell^1+L_1,\ell^2+L_1} $\\
    $\psi_{\vec{\ell}} = f(\vec{\ell})\psi_{-\ell^2+S_2^1,-\ell^1-S_2^2}$\\ $f(\vec{\ell})=(-1)^{\ell^1\ell^2+\ell^2+\ell^1}$} & 0 & 0 & 0& 2 & $(1, \sR\Gamma)$\\
    \hline 
    $(1,T_1\R)$ & \makecell{$\psi_{\vec{\ell}}=\psi_{\ell^1+L_1,\ell^2}$,\\$\psi_{\vec{\ell}}=f(\vec{\ell})\psi_{-\ell^1+S_1^1,\ell^2+S_1^1}$\\
    $f(\vec{\ell})=(-1)^{\ell_1+\ell_2}$} & 0& 1 &0 & 0 & $(1, {\sR}e^{i\frac{\pi}{2}\sJ})$\\
    \hline
    $(T_1T_2,\C\R)$ & \makecell{$\psi_{\vec{\ell}} = (-1)^{\ell_2} \psi_{\ell^1+L_1,\ell^2+L_1} $\\
    $\psi_{\vec{\ell}} = f(\vec{\ell})\psi_{-\ell^2+S_2^1,-\ell^1-S_2^2}$\\ $f(\vec{\ell})=(-1)^{\ell^1\ell^2+\ell^2+\ell^1}$} & 1 & 0 & 0 & 0 & $(e^{i\frac{\pi}{2}\sJ}, {\sR}\Gamma)$\\
    \hline 
    $(1,T_2\C\R)$ & \makecell{$\psi_{\vec{\ell}} =  \psi_{\ell^1+L_1,\ell^2+L_1} $\\
    $\psi_{\vec{\ell}} = f(\vec{\ell})\psi_{-\ell^2+S_2^1,-\ell^1-S_2^2}$\\ $f(\vec{\ell})=(-1)^{\ell^1\ell^2+\ell^2+\ell^1}$} & 0 & 0  & 1& 0 & $(1\,, \sR e^{i\frac{\pi}{4} J})$\\
    \hline 
    $(1,T_1\C\R)$ & \makecell{$\psi_{\vec{\ell}} =  \psi_{\ell^1+L_1,\ell^2+L_1} $\\
    $\psi_{\vec{\ell}} = f(\vec{\ell})\psi_{-\ell^2+S_2^1,-\ell^1-S_2^2}$\\ $f(\vec{\ell})=(-1)^{\ell^1\ell^2+\ell^2}$} & 0 & 1  & 0& 0 & $(1\,, \sR e^{i\frac{3\pi}{4} J})$\\
    \hline 
    $((-1)^F,g_2)$ & $\dots$ & $\dots$  & $\dots$  & $\dots$  & 0 & $((-1)^F,\bar k_2)$\\
    \hline 
    \end{tabular}
    \caption{Sixteen inequivalent lattice twists (seven from the last row) on the Klein bottle and their continuum counterparts. For each lattice twist, we list the corresponding boundary conditions labeled by $L_1 $ and $S_1^i $ modulo 2, and the number of real fermion zero modes. The final row records the insertion of a $(-1)^F$ defect along the untwisted cycle 1.  On the lattice, it turns $(1,g_2)$ to $((-1)^F,g_2)$, and in the continuum, it turns $(1,\bar k_2)$ to $((-1)^F,\bar k_2)$.}
    \label{tab:KBtwist}
\end{table}

\section{Anomalies involving $\Omega$}
\label{app:Omegaanomaly}
Here, following \cite{Seiberg:2025zqx}, we reinterpret the projective phases involving the anti-unitary symmetry $\Omega$ as the time-reversal anomaly of a $0+1$d quantum–mechanical system. This is related to the lattice realization of the Smith homomorphism \cite{gilkey1989geometry,Tachikawa:2018njr,Kapustin:2014dxa,Hason:2020yqf,Cordova:2019wpi}.

We place the theory on a rectangular torus and begin with even $L_1,L_2$. The action of $\Omega$ has four fixed points on the lattice,
\ie
\Omega\,\psi_{0,0}\,\Omega^{-1} &= \psi_{0,0}\,, \qquad 
\Omega\,\psi_{0,\frac{L_2}{2}}\,\Omega^{-1} = \psi_{0,\frac{L_2}{2}}\,,\\
\Omega\,\psi_{\frac{L_1}{2},0}\,\Omega^{-1} &= \psi_{\frac{L_1}{2},0}\,, \qquad
\Omega\,\psi_{\frac{L_1}{2},\frac{L_2}{2}}\,\Omega^{-1} = \psi_{\frac{L_1}{2},\frac{L_2}{2}}\,.
\fe
All other sites are paired by $\Omega$ and lead to the eigenmodes
\ie
\psi^{(\pm)}_{\vec\ell}\equiv \psi_{\vec\ell}\pm\psi_{-\vec\ell}
\,,\qquad
\Omega\,\psi^{(\pm)}_{\vec\ell}\,\Omega^{-1}=\pm\,\psi^{(\pm)}_{\vec\ell}\,,
\fe
which can be gapped by an $\Omega$-invariant mass. Hence, only the four fermions at the fixed points remain unpaired and contribute to the anomaly. Equivalently, $\Omega$ acts as time-reversal on a $0{+}1$d system of four fermions, and the projective phases have order $2$.

Next, keep $L_2$ even but take $L_1$ odd. The number of fixed points drops to two,
\ie
\Omega\,\psi_{0,0}\,\Omega^{-1} &= \psi_{0,0}\,,\qquad
\Omega\,\psi_{0,\frac{L_2}{2}}\,\Omega^{-1} = \psi_{0,\frac{L_2}{2}}\,,
\fe
while all the other sites are again paired, and the fermions there can be gapped as above. Thus $\Omega$ acts as a time-reversal on two unpaired fermions, and the projective phases have order 4.

Finally, consider the Klein bottle obtained by the twisted identifications
\ie
\psi_{\ell^1,\ell^2}=(-1)^{\ell^2}\psi_{\ell^1+L_1,\ell^2}\,,\qquad
\psi_{\ell^1,\ell^2}=(-1)^{\ell^1}\psi_{-\ell^1,\ell^2+L_2}\,,
\fe
with odd $L_{1,2}$. There is a single fixed point at $(0,0)$. All the other modes are paired and can be gapped. In this case, $\Omega$ acts as a time-reversal operation on the single unpaired fermion, and the projective phases have order 8.

In summary, by pairing all the $\Omega$-related sites and reducing to an effective quantum mechanics of the unpaired fermions at the fixed points, we map the anomaly of $\Omega$ in 2+1d to the time-reversal anomaly in quantum mechanics. (All other cases in Section \ref{sec:latticeanomaly} can be analyzed using the same logic.)

\bibliographystyle{JHEP}
\bibliography{references}

\end{document}